\documentclass[reprint,amsmath,amssymb,prl,superscriptaddress]{revtex4-1}

\usepackage{graphicx}
\usepackage{bm}
\usepackage{amsmath,amssymb}
\usepackage{xspace}
\usepackage{braket}
\usepackage{comment}
\usepackage{mathrsfs}
\usepackage[left]{lineno}

\AtBeginDocument{\let\oldcontentsline\contentsline}
\newcommand{\notoccontentsline}[4]{\oldcontentsline{}{}{}{}}
\newcommand{\droptocpage}{\addtocontents{toc}{\let\protect\contentsline\protect\notoccontentsline}}
\newcommand{\incltocpage}{\addtocontents{toc}{\let\protect\contentsline\protect\oldcontentsline}}

\begin{document}
	\title{Optical mode conversion via spatiotemporally modulated atomic susceptibility}
	\author{Claire Baum}
	\author{Matt Jaffe}
	\author{Lukas Palm}
	\author{Aishwarya Kumar}
	\affiliation{James Franck Institute and Department of Physics, University of Chicago, Chicago, IL 60637, USA}
	\author{Jonathan Simon}
	\affiliation{The Department of Physics, The James Franck Institute, and The Pritzker School of Molecular Engineering, The University of Chicago, Chicago, IL}
	\affiliation{The Department of Physics, Stanford University, Stanford, CA}
	\affiliation{The Department of Applied Physics, Stanford University, Stanford, CA}
	
	\date{\today}
	\begin{abstract} 
		Light is an excellent medium for both classical~\cite{Agrell2016RoadmapCommunications} and quantum information~\cite{gisin2007quantum} transmission due to its speed, manipulability, and abundant degrees of freedom into which to encode information~\cite{Winzer2014MakingReality}. Recently, space-division multiplexing~\cite{Richardson2013Space-divisionFibres,Winzer2012OpticalWDM, Winzer2013SpatialScaling, Xavier2020QuantumFibres, Puttnam2021Space-divisionCommunications, Su2021PerspectiveMultiplexing,Winzer2014MakingReality, Su2021PerspectiveMultiplexing, Xia2014Space-divisionCommunication, Willner2019UsingSorter} has gained attention as a means to substantially increase the rate of information transfer~\cite{Patel2014QuantumNetworks, Bozinovic2013Terabit-scaleFibers, Sakaguchi2012SpaceFiber, Wang2012TerabitMultiplexing} by utilizing sets of infinite-dimensional propagation eigenmodes such as the Laguerre-Gaussian ‘donut’ modes~\cite{willner2015optical, Molina-Terriza2007TwistedPhotons, erhard2018twisted}. Encoding in these high-dimensional spaces necessitates devices capable of manipulating photonic degrees of freedom with high efficiency. In this work, we demonstrate controlling the optical susceptibility of an atomic sample can be used as powerful tool for manipulating the degrees of freedom of light that passes through the sample. Utilizing this tool, we demonstrate photonic mode conversion between two Laguerre-Gaussian modes of a twisted optical cavity with high efficiency. We spatiotemporally modulate~\cite{Clark2019InteractingPolaritons} the optical susceptibility of an atomic sample that sits at the cavity waist using an auxiliary Stark-shifting beam, in effect creating a mode-coupling optic that converts modes of orbital angular momentum $l=3\rightarrow l=0$. The internal conversion efficiency saturates near unity as a function of the atom number and modulation beam intensity, finding application in topological few-body state preparation~\cite{Ivanov2018AdiabaticPolaritons}, quantum communication~\cite{kimble2008quantum, willner2015optical}, and potential development as a flexible tabletop device.
	\end{abstract}
	
	\maketitle
	
	Efficient control over photonic degrees of freedom, including frequency, polarization, and spatial mode, has widespread applications in information and communication. Put simply: the more degrees of freedom one can manipulate, the more information one can encode in a single channel of light. This idea is utilized regularly in both classical and quantum communication, where light has been multiplexed in arrival time~\cite{baharudin2013review, sangdeh2019overview, Winzer2012OpticalWDM}, frequency~\cite{ishio1984review, baharudin2013review, sangdeh2019overview, Winzer2012OpticalWDM}, polarization~\cite{Winzer2012OpticalWDM}, quadrature~\cite{Winzer2012OpticalWDM}, and most recently space~\cite{Richardson2013Space-divisionFibres,Winzer2012OpticalWDM, Winzer2013SpatialScaling, Xavier2020QuantumFibres, Puttnam2021Space-divisionCommunications, Su2021PerspectiveMultiplexing,Winzer2014MakingReality, Su2021PerspectiveMultiplexing, Xia2014Space-divisionCommunication, Willner2019UsingSorter} to substantially increase information transfer over a fiber~\cite{Patel2014QuantumNetworks, Bozinovic2013Terabit-scaleFibers, Sakaguchi2012SpaceFiber} and free-space link~\cite{Wang2012TerabitMultiplexing}. Spatial information may be conveniently encoded within families of propagation eigenmodes; the Hermite-Gaussian (HG) and Laguerre-Gaussian (LG) families are appealing for their orthogonality and infinite-dimensionality, supporting the exploration of higher-dimensional Hilbert spaces for quantum computing~\cite{ralph2007efficient, chen2017realization}, formation of orbital angular momentum qudits~\cite{chen2017realization, Molina-Terriza2007TwistedPhotons, Goyal2013TeleportingScissors, Xavier2020QuantumFibres, Garcia-Escartin2008QuantumLight, cozzolino2019high, erhard2018twisted}, improved quantum key distribution~\cite{Molina-Terriza2007TwistedPhotons, Mirhosseini2015High-dimensionalLight, Xavier2020QuantumFibres, erhard2018twisted, willner2015optical, Mair2001EntanglementPhotons, Krenn2014GenerationSystem}, lower-crosstalk quantum communication~\cite{ren2015free, tariq2021orbital,willner2015optical}, and distribution of quantum information to multiple users in a quantum network~\cite{Garcia-Escartin2008QuantumLight}.
	
	\begin{figure*}
		\centering
		\includegraphics[width=1.0\textwidth]{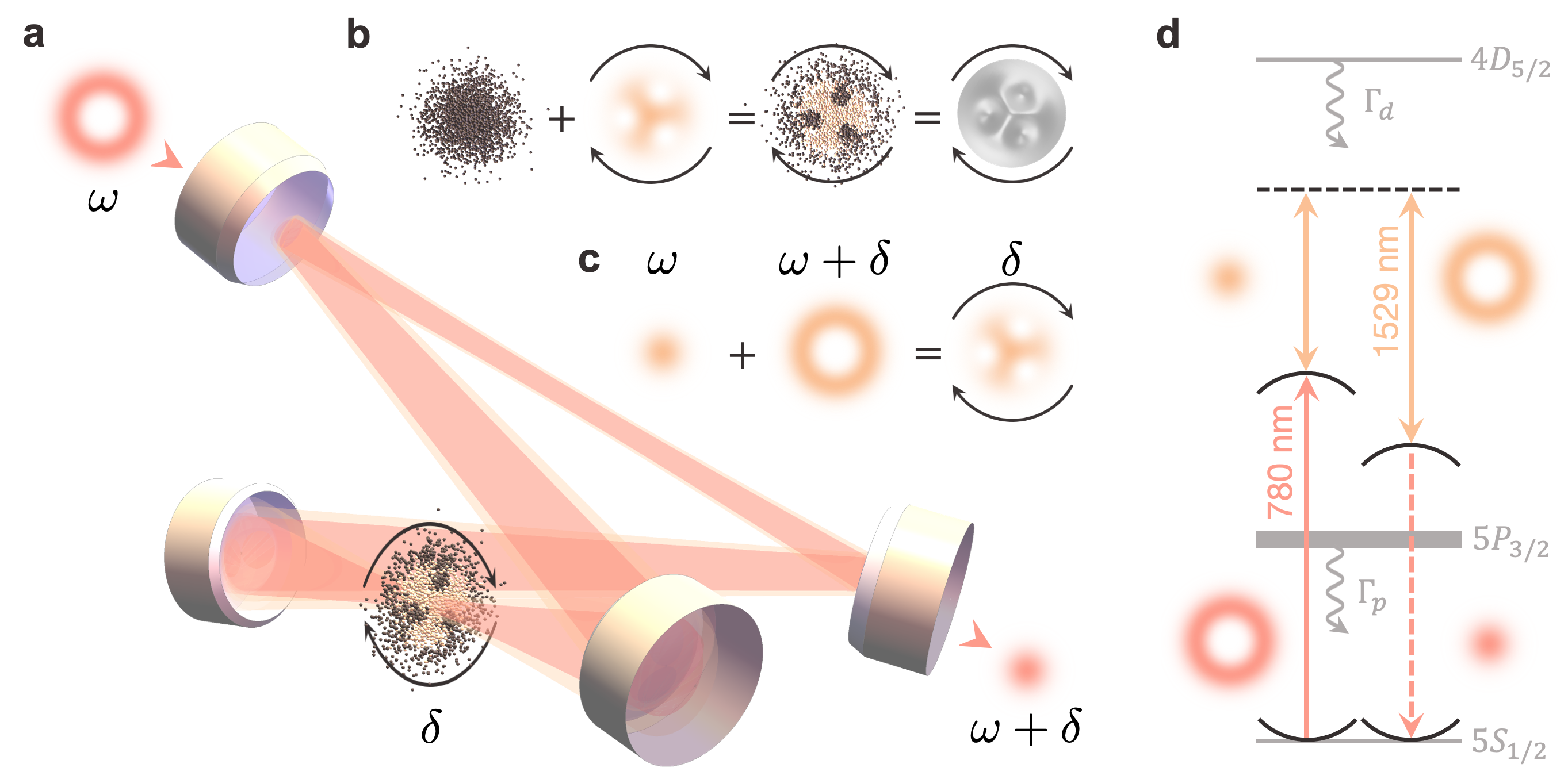} 
		\caption{
			\textbf{Modulated atomic samples as sculptable optics.}
			This work demonstrates the conversion of photons between two Laguerre-Gaussian modes of orbital angular momenta $l=0$ and $l=3$. These modes are the non-degenerate eigenmodes of the twisted cavity depicted in \textbf{a}, which hosts $780$~nm probe modes (red) and slightly larger, copropagating $1529$~nm modulation modes (orange). We inject $l=3$ probe photons which are converted to $l=0$ via coupling to an atomic sample of $^{87}$Rb atoms at the waist of the cavity. \textbf{b,} The optical susceptibility of this sample is modulated in space and time by the $1529$~nm modulation beam, effectively sculpting a rotating, mode-coupling optic from the atomic cloud with a spatiotemporally-varying refractive index. This coupling arises from the spatial profile used modulate the atomic sample, which is comprised of both $l=0$ and $l=3$ modes as illustrated in \textbf{c}. When an $l=0$ mode of frequency $\omega$ is spatially overlapped with an $l=3$ mode of frequency $\omega+\delta$, the resulting interference profile is a three-holed pattern that rotates at the frequency difference between the two modes. When the atomic sample is illumated with this rotating profile, the optical susceptibility of the sample is modulated according to the profile of the modulation beam, effectively sculpting the stationary sample into a rotating, three-fold symmetric optic. The relevant atomic levels for this mode conversion process are illustrated in \textbf{d}. We inject $780$~nm, $l=3$ probe photons in the dispersive regime, $130$~MHz detuned from the $5S_{1/2}\leftrightarrow5P_{3/2}$ atomic resonance. These photons are coupled by the far-detuned $1529$~nm modulation beam to the $l=0$ cavity mode at $780$~nm. If mode conversion is successful, $l=0$ photons will emerge from the cavity at a frequency $65$~MHz lower than that of the injected $l=3$ photons as a result of the frequency difference between non-degenerate cavity eigenmodes. The optical susceptibility of the atomic cloud is modulated via the time-varying, spatially-dependent optical Stark shift of the $5P_{3/2}$ energy, which periodically shifts the $5S_{1/2}\leftrightarrow5P_{3/2}$ atomic resonance further from the cavity resonances.
			\label{fig:Intro}}
	\end{figure*}
	
	High-dimensional optical information encoding requires the ability to \emph{manipulate} the various photonic degrees of freedom through ‘mode conversion’~\cite{Shen2022ModeBeams, Liang2019ControllablePhase, fontaine2019laguerre, zhou2018hermite, beijersbergen1993astigmatic, yao2011orbital, Danaci2016All-opticalMixing, Nie2016MultichannelCommunication, Shen2022OAMChip, Pires2019OpticalMixing, Willner2019UsingSorter}. Frequency and polarization mode conversion can be achieved quite flexibly at near-unit efficiency using electro-optic modulators~\cite{cumming1957serrodyne} and waveplates. However, efficient spatial mode conversion is more challenging. In general, spatial mode conversion requires a spatially-dependent phase and amplitude modification of a photon's electric field. While phase can be modified losslessly by a phase-imprinting device, amplitude modification occurs only through propagation or discarding amplitude via a physical barrier, limiting the efficiency with which spatial mode conversion can occur. For instance, devices such as spatial light modulators, digital micromirror devices, vortex plates, and liquid crystal q-plates~\cite{Piccirillo2009LightCharge, Piccirillo2011EfficientLinks, Karimi2009EfficientQ-plates, slussarenko2013liquid, karimi2009light} are excellent devices for generating modes with orbital angular momentum (OAM) by imprinting incident light with a spiral phase. While the resulting mode has the correct phase winding to be purely LG, its amplitude distribution does not. Rather, the resulting mode can be expressed as an expansion of the LG radial modes for a given OAM, illustrating that a phase imprint alone is insufficient for highly efficient spatial mode conversion to a \emph{single} LG mode~\cite{wei2019generating, willner2015optical}. Thus, mode-converting devices have been designed to modify light in environments such as waveguides, cavities, and photonic crystals that limit the occupiable spatial modes to enhance conversion to a single target mode. Among these devices are a HG$\leftrightarrow$LG mode converter using an astigmatic microcavity~\cite{Nakagawa2020LaguerreGaussianMicrocavity}, an arbitrary HG mode-order converter utilizing the impedance mismatches between coupled Fabry-P\'erot resonators~\cite{stone2021optical}, design-by-specification converters based on computational methods~\cite{lu2013nanophotonic}, and an assortment of silicon photonic converters that harness refractive index variation to smoothly modify a propagating spatial mode~\cite{Li2018MultimodePhotonics, Wang2019CompactStructure, Karabchevsky2019SpatialWaveguides,Wajih2019ASlots, Wang2020Ultra-compactSlots, Zhang2020On-chipCrosstalk, Zhu2021Silicon-BasedMetasurface,Miller2012Ultra-compactEffect, Huang2006AnConverter,Dai2012ModeWaveguides, dai2013silicon, Frandsen2014TopologyMaterial, Chen2005WaveguideCrystals}.
	
	In this paper, we explore a new method in which spatial and frequency mode conversion occur simultaneously in a single system with high efficiency. In effect, we create a rapidly sculptable, rotating optic inside of an optical cavity that converts photons between cavity modes. In practice, we modulate~\cite{Clark2019InteractingPolaritons}, in both space and time, the optical susceptibility of a stationary atomic sample at the waist of a twisted optical cavity using a strong auxiliary beam, inducing a coupling between cavity modes. This auxiliary beam Stark shifts the energy levels of the atomic sample to create a spatiotemporally-varying optical susceptibility across the atomic sample akin to a rotating optic. Photons that are incident on the atomic sample accrue a position-dependent phase that couples the incident mode to other modes of the cavity, which enables repeated light-atom interactions and preferentially enhances the emission of light into supported, resonant spatial modes. We measure the efficiency of this conversion process for increasing atom number and modulation beam intensity. We find a parameter regime in which the internal conversion efficiency saturates near unity.

	\begin{figure*}
		\centering
		\includegraphics[width=1.0\textwidth]{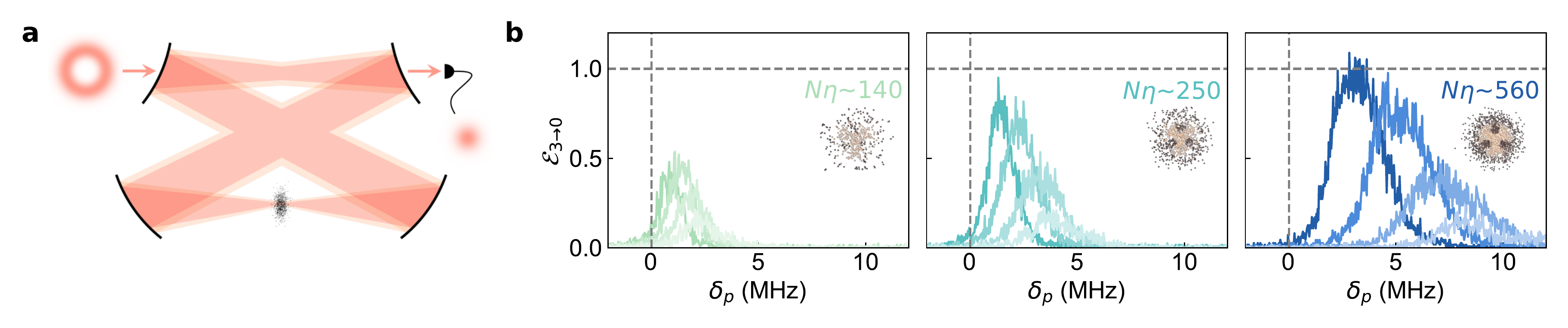}
		\caption{
			\textbf{Conversion in the cavity spectra.}
			We inject $l=3$ probe photons into the twisted cavity and read out only on $l=0$ using a single mode fiber as illustrated in \textbf{a}. For various combinations of $\Omega$ and $N\eta$, corresponding with the modulation beam intensity and atom number, respectively, we observe the $l=0$ transmission spectrum by scanning the frequency of the probe laser frequency, $\delta_p$, about the bare $l=3$ transmission frequency ($\delta_p=0$ MHz). In \textbf{b}, we plot the $l=0$ transmission spectrum (normalized as an internal conversion efficiency) at $\Omega/(2\pi)=(0.7, 1.3, 2.1, 3.5)$ GHz (light to dark) for each value of $N\eta=(140, 250, 560)$. The $l=3\rightarrow l=0$ conversion efficiency, $\mathcal{E}_{3\rightarrow0}$, increases for increasing $\Omega$ and $N\eta$, reaching near unity for the highest values of $\Omega$ and $N\eta$. Intuitively, $\mathcal{E}_{3\rightarrow0}$ should increase for increasing modulation beam intensity and atom number, akin to increasing the refractive index variation and density of our effective, intracavity optic from zero. The $l=0$ transmission curves shift toward lower frequencies with increasing $\Omega$ due to the increasing Stark shift of the $5P_{3/2}$ resonance with higher modulation beam intensity, which lessens the dispersive shift of the $l=0$ transmission curves away from the bare $l=3$ transmission frequency.
			\label{fig:ConvSpectra}}
	\end{figure*}
	
	We demonstrate conversion between LG modes of orbital angular momenta $l=3 \rightarrow l=0$ (i.e. LG$_{30}\rightarrow$LG$_{00}$). Our optical cavity is a four-mirror twisted cavity, meaning one mirror lies outside of the plane formed by the remaining three~\cite{Schine2016SyntheticPhotons}. As the eigenmodes of this cavity are non-degenerate LG modes, cavity photons require a change in both their spatial and frequency degrees of freedom to undergo mode conversion. This change can be accomplished by passage through the an atomic sample whose optical susceptibility varies in time and space. Provided the variation occurs at the frequency difference between $l=0$ and $l=3$ and imprints a phase on $l=0 (3)$ such that the resulting spatial mode has non-zero overlap with $l=3 (0)$, a coupling will be engineered between the $l=0$ and $l=3$ cavity modes.
	
	Fig.~\ref{fig:Intro}a illustrates our mode conversion scheme. A $^{87}$Rb atomic sample resides at the waist of our twisted optical cavity, which hosts modes at $780$~nm (near the $5\textrm{S}_{1/2}\leftrightarrow 5\textrm{P}_{3/2}$ transition of $^{87}$Rb) and at $1529$~nm (near the $5\textrm{P}_{3/2}\leftrightarrow 4\textrm{D}_{5/2}$ transition of $^{87}$Rb). The $5\textrm{P}_{3/2}\leftrightarrow 4\textrm{D}_{5/2}$ transition of the atomic sample is energetically modulated by a time-varying, spatially-dependent optical Stark shift generated by an auxiliary `modulation' beam at $1529$~nm whose intensity distribution is illustrated in Fig.~\ref{fig:Intro}b. This pattern is achieved by overlapping $1529$~nm $l=0$ and $l=3$ modes, forming an intensity profile with three ‘holes' that rotates at the frequency difference ($\approx 65$~MHz) between the modes. Illuminating the atomic sample with this profile changes the resonance condition of individual atoms with intracavity $780$~nm photons, creating a spatiotemporally-varying optical susceptibility across the sample that adopts the modulation beam profile (Fig.~\ref{fig:Intro}c). As the modulation beam profile is comprised of both the $l=0$ and $l=3$ modes, a coupling is engineered between the $l=0$ and $l=3$ modes at $780$~nm. Note that the atomic sample is stationary whereas the modulation profile rotates, enabling far faster temporal modulation of incident probe light than that which can be achieved by a real, rotating optic. We utilize the atomic level scheme illustrated in Fig.~\ref{fig:Intro}d, which may be understood as a near-resonant four-wave mixing process.
	
	We begin our experimental sequence by transporting a sample of laser-cooled $^{87}$Rb into the waist of our twisted optical cavity from a magneto-optical trap. The modulation beam and weak probe beam co-propagate through the cavity and illuminate the atomic sample for a probe time of $10$~ms. Probe photons are injected into the $l=3$ cavity eigenmode. These photons pass through the modulated atomic sample and the resulting photons are collected on the cavity output during the probe time. See SI. Fig.~\ref{fig:ExpSetup} for additional details about the experimental setup.
	
	\begin{figure*}
		\centering
		\includegraphics[width=1.0\textwidth]{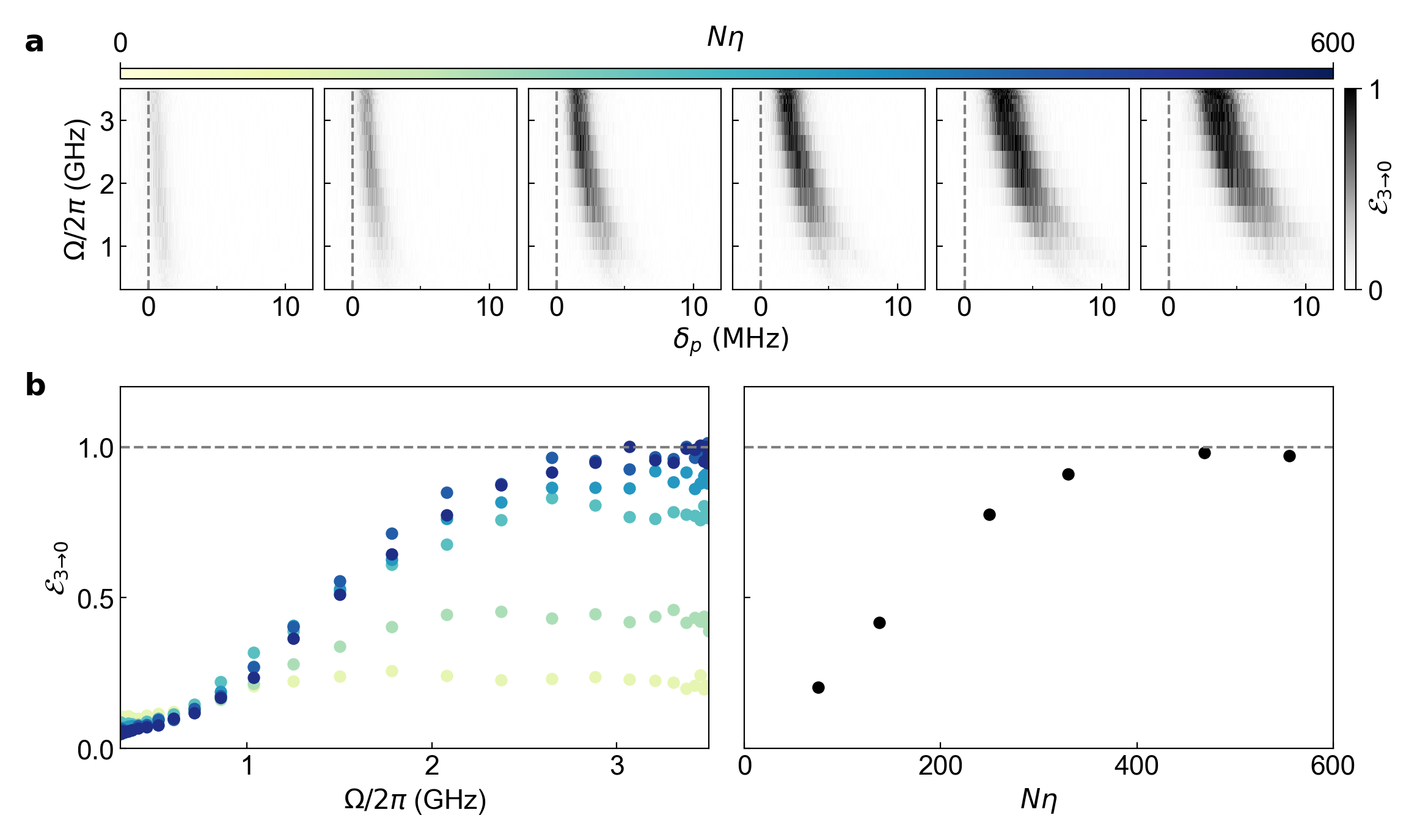}
		\caption{
			\textbf{Saturation in conversion efficiency.} 
			We further examine $l=3\rightarrow l=0$ conversion in two-dimensional sweeps over $\delta_p$ and $\Omega$ for additional values of $N\eta$. In \textbf{a}, $\mathcal{E}_{3\rightarrow0}$ increases for increasing $\Omega$ and $N\eta$. Plotting the numerical values of $\mathcal{E}_{3\rightarrow0}$ in \textbf{b} at each slice of $\Omega$ for all $N\eta$ \textbf{(left)} and for each $N\eta$ at maximum $\Omega$ \textbf{(right)} elucidates both the attainment and \emph{saturation} of conversion near $\mathcal{E}_{3\rightarrow0}$=1. This internal efficiency corresponds to a maximum external efficiency of 25\% due to the double-ended nature of our cavity. In general, light is fully transmitted through a double-ended cavity when the reflected light cancels with the light that leaks out of the cavity. This idea assumes the two cavity ends, or mirrors, have equal transmission coefficients and light drives the cavity from one side. Even though the two ends of our twisted cavity have equal transmission coefficients, the conversion of light from the injected mode to another mode acts as loss which breaks the cavity impedance matching that enables full transmission. See SI~\ref{SI:Calib} and ~\ref{SI:ImMatch} for more details. Points are larger than their error bars of one standard deviation.
			\label{fig:ConvSpatialFreq}}
	\end{figure*}
	
	We search for $l=3\rightarrow l=0$ mode conversion for several different combinations of modulation beam intensity and atom number by collecting only $l=0$ light from the cavity using a single mode fiber as a filter (Fig.~\ref{fig:ConvSpectra}a). For each of these combinations, we scan the frequency of the probe beam about a point in the dispersive regime, where the $l=0$ and $l=3$ cavity resonances are detuned from the atomic $5P_{3/2}$ state as illustrated in Fig.~\ref{fig:Intro}d. This scan generates the $l=0$ spectra in Fig.~\ref{fig:ConvSpectra}b. We observe an increase in the $l=3\rightarrow l=0$ internal conversion efficiency, $\mathcal{E}_{3\rightarrow0}$, for increasing $\Omega$ and $N\eta$, in effect the modulation beam intensity and resonant optical density, respectively. See SI~\ref{SI:Calib} for additional details about $\mathcal{E}_{3\rightarrow0}$. Here, $N\eta$ is the collective cooperativity~\cite{tanji2011interaction} where $N$ is the atom number and $\eta$ is the single atom cooperativity. This quantity can be generally interpreted as the number of times a photon is lensed by the atomic sample before it leaks out of the cavity. See Methods and SI~\ref{SI:DefOmega} for additional details about $N\eta$ and $\Omega$, respectively. As $\Omega$ increases, we observe the $l=0$ cavity transmissions collapse leftward toward the location of the bare $l=3$ transmission at $\delta_p=0$. This behavior is a result of the $5P_{3/2}$ state energetically shifting away from the $l=0$ and $l=3$ cavity resonances at higher modulation beam intensities, reducing the dispersive shift of the resonances.
	
	To verify photons are indeed converted into the $l=0$ mode of the cavity, we perform a spatial and frequency analysis of the cavity output. In principle, the modulated atomic sample induces a coupling between the $l=3$ mode and many other spatial modes. However, with the exception of the $l=6$ mode, these modes are Purcell suppressed because they are non-resonant. Despite a potential $3\leftrightarrow 6$ coupling, we do not observe $l=6$ light on the cavity output, likely because the $l=6$ mode is further detuned from the $5\textrm{S}_{1/2}\leftrightarrow 5\textrm{P}_{3/2}$ atomic resonance compared to $l=3$ and $l=0$ modes (see SI~\ref{SI:actual0}). Thus, in general, the non-degenerate mode structure of the cavity improves the isolation of a target mode by frequency discrimination. 
	
	The increase of $\mathcal{E}_{3\rightarrow0}$ with $\Omega$ and $N\eta$ can be interpreted intuitively in the context of sculpting an effective optic from the atomic sample. For $\Omega=0$, there is no modulation of the atomic sample. Probe photons pass through an effective optic that imparts an almost completely flat phase, providing essentially no coupling between the $l=3$ and $l=0$ modes. For $N\eta=0$, no atoms are present; there is no effective optic. Thus, $\mathcal{E}_{3\rightarrow0}$ regardless of $\Omega$. For $\Omega>0$ and $N\eta>0$, we begin to observe $l=3$ to $l=0$ conversion as the effective optic acquires density and a spatially-dependent optical susceptibility.
	
	Fig.~\ref{fig:ConvSpatialFreq} is a more in-depth investigation of $\mathcal{E}_{3\rightarrow0}$ as a function of $\Omega$ and $N\eta$. $\mathcal{E}_{3\rightarrow0}$ increases for increasing $\Omega$ and $N\eta$ and \emph{saturates} near unity. In a double-ended cavity like ours, where light can leak out one of two cavity mirrors, $\mathcal{E}_{3\rightarrow0}$=1 corresponds to a maximum \emph{external} efficiency of 25\% for lossless mirrors. For a general double-ended cavity comprised of two equally-transmissive cavity mirrors, incident light can be fully transmitted as the cavity is impedance matched. If a mode-converting element is placed within the cavity, this impedance matching condition is broken, limiting the amount of light, both converted and unconverted, that exits the cavity through the output mirror. In a single-ended cavity, the maximum external efficiency increases to 100\% (see SI~\ref{SI:ImMatch}).
	
	We have demonstrated a highly efficient method to simultaneously manipulate photonic degrees of freedom by spatiotemporally modulating the optical susceptibility of an atomic sample. In our twisted optical cavity, we observe $l=3\rightarrow l=0$ conversion at an internal efficiency near unity. Extending this method to a low loss, single-ended cavity will provide conversion near 100\% efficiency for both internal and external efficiencies. This method is additionally extendable to other atomic species, arbitrary cavity geometries, different propagation eigenmodes, polarization conversion (see Methods), and the coherent conversion of single photons~\cite{kimble2008quantum}. Mode conversion via optical susceptibility modulation might also find applications in quantum state preparation, quantum information, and development as a tabletop device. One might use this method to grow topological few-body states of light by controllably adding orbital angular momentum to intracavity photons~\cite{Ivanov2018AdiabaticPolaritons}, convert within mode pairs for mode-division multiplexed transmission~\cite{willner2015optical}, or create a miniaturized device based on intracavity electro-optic elements whose refractive indices are modulated in space and time.

	\droptocpage
	
	\section{Acknowledgements}
	We acknowledge conversations with M. Fleischhauer. This work was supported by AFOSR Grant FA9550-18-1-0317 and AFOSR MURI FA9550-19-1-0399. C.B. acknowledges support from the NSF Graduate Research Fellowships Program (GRFP).
	
	\section{Author Contributions}
	C.B., M.J., and J.S. designed the experiment. C.B. and L.P. built the experiment. C.B. collected and analyzed the data. C.B., M.J., A.K., and J.S. contributed to the theoretical model. C.B. wrote, and all authors contributed to, this manuscript.
	
	\section{Author Information}
	The authors declare no competing financial interests. Correspondence and requests for materials should be addressed to J.S. (jonsimon@stanford.edu). 
	
	\section{Data Availability}
	The experimental data presented in this manuscript are available from the corresponding author upon request.
	
	\bibliographystyle{naturemag}
	\bibliography{AOMOAM_V2}

\begin{thebibliography}{10}
\expandafter\ifx\csname url\endcsname\relax
  \def\url#1{\texttt{#1}}\fi
\expandafter\ifx\csname urlprefix\endcsname\relax\def\urlprefix{URL }\fi
\providecommand{\bibinfo}[2]{#2}
\providecommand{\eprint}[2][]{\url{#2}}

\bibitem{Agrell2016RoadmapCommunications}
\bibinfo{author}{Agrell, E.} \emph{et~al.}
\newblock {Roadmap of optical communications}.
\newblock \emph{\bibinfo{journal}{Journal of Optics}}
  \textbf{\bibinfo{volume}{18}}, \bibinfo{pages}{063002}
  (\bibinfo{year}{2016}).

\bibitem{gisin2007quantum}
\bibinfo{author}{Gisin, N.} \& \bibinfo{author}{Thew, R.}
\newblock Quantum communication.
\newblock \emph{\bibinfo{journal}{Nature photonics}}
  \textbf{\bibinfo{volume}{1}}, \bibinfo{pages}{165--171}
  (\bibinfo{year}{2007}).

\bibitem{Winzer2014MakingReality}
\bibinfo{author}{Winzer, P.~J.}
\newblock {Making spatial multiplexing a reality}.
\newblock \emph{\bibinfo{journal}{Nature Photonics 2014 8:5}}
  \textbf{\bibinfo{volume}{8}}, \bibinfo{pages}{345--348}
  (\bibinfo{year}{2014}).

\bibitem{Richardson2013Space-divisionFibres}
\bibinfo{author}{Richardson, D.~J.}, \bibinfo{author}{Fini, J.~M.} \&
  \bibinfo{author}{Nelson, L.~E.}
\newblock {Space-division multiplexing in optical fibres}.
\newblock \emph{\bibinfo{journal}{Nature Photonics 2013 7:5}}
  \textbf{\bibinfo{volume}{7}}, \bibinfo{pages}{354--362}
  (\bibinfo{year}{2013}).

\bibitem{Winzer2012OpticalWDM}
\bibinfo{author}{Winzer, P.~J.}
\newblock {Optical Networking Beyond WDM}.
\newblock \emph{\bibinfo{journal}{IEEE Photonics Journal}}
  \textbf{\bibinfo{volume}{2}}, \bibinfo{pages}{647 -- 651}
  (\bibinfo{year}{2012}).

\bibitem{Winzer2013SpatialScaling}
\bibinfo{author}{Winzer, P.~J.}
\newblock {Spatial multiplexing: The next frontier in network capacity
  scaling}.
\newblock \emph{\bibinfo{journal}{IET Conference Publications}}
  \textbf{\bibinfo{volume}{2013}}, \bibinfo{pages}{372--374}
  (\bibinfo{year}{2013}).

\bibitem{Xavier2020QuantumFibres}
\bibinfo{author}{Xavier, G.~B.} \& \bibinfo{author}{Lima, G.}
\newblock {Quantum information processing with space-division multiplexing
  optical fibres}.
\newblock \emph{\bibinfo{journal}{Communications Physics 2020 3:1}}
  \textbf{\bibinfo{volume}{3}}, \bibinfo{pages}{1--11} (\bibinfo{year}{2020}).

\bibitem{Puttnam2021Space-divisionCommunications}
\bibinfo{author}{Puttnam, B.~J.}, \bibinfo{author}{Rademacher, G.} \&
  \bibinfo{author}{Lu{\'{i}}s, R.~S.}
\newblock {Space-division multiplexing for optical fiber communications}.
\newblock \emph{\bibinfo{journal}{Optica, Vol. 8, Issue 9, pp. 1186-1203}}
  \textbf{\bibinfo{volume}{8}}, \bibinfo{pages}{1186--1203}
  (\bibinfo{year}{2021}).

\bibitem{Su2021PerspectiveMultiplexing}
\bibinfo{author}{Su, Y.}, \bibinfo{author}{He, Y.}, \bibinfo{author}{Chen, H.},
  \bibinfo{author}{Li, X.} \& \bibinfo{author}{Li, G.}
\newblock {Perspective on mode-division multiplexing}.
\newblock \emph{\bibinfo{journal}{Applied Physics Letters}}
  \textbf{\bibinfo{volume}{118}}, \bibinfo{pages}{200502}
  (\bibinfo{year}{2021}).

\bibitem{Xia2014Space-divisionCommunication}
\bibinfo{author}{Xia, C.}, \bibinfo{author}{Li, G.}, \bibinfo{author}{Bai, N.}
  \& \bibinfo{author}{Zhao, N.}
\newblock {Space-division multiplexing: the next frontier in optical
  communication}.
\newblock \emph{\bibinfo{journal}{Advances in Optics and Photonics, Vol. 6,
  Issue 4, pp. 413-487}} \textbf{\bibinfo{volume}{6}},
  \bibinfo{pages}{413--487} (\bibinfo{year}{2014}).

\bibitem{Willner2019UsingSorter}
\bibinfo{author}{Willner, A.~E.} \emph{et~al.}
\newblock {Using all transverse degrees of freedom in quantum communications
  based on a generic mode sorter}.
\newblock \emph{\bibinfo{journal}{Optics Express, Vol. 27, Issue 7, pp.
  10383-10394}} \textbf{\bibinfo{volume}{27}}, \bibinfo{pages}{10383--10394}
  (\bibinfo{year}{2019}).

\bibitem{Patel2014QuantumNetworks}
\bibinfo{author}{Patel, K.~A.} \emph{et~al.}
\newblock {Quantum key distribution for 10 Gb/s dense wavelength division
  multiplexing networks}.
\newblock \emph{\bibinfo{journal}{Applied Physics Letters}}
  \textbf{\bibinfo{volume}{104}}, \bibinfo{pages}{051123}
  (\bibinfo{year}{2014}).

\bibitem{Bozinovic2013Terabit-scaleFibers}
\bibinfo{author}{Bozinovic, N.} \emph{et~al.}
\newblock {Terabit-scale orbital angular momentum mode division multiplexing in
  fibers}.
\newblock \emph{\bibinfo{journal}{Science}} \textbf{\bibinfo{volume}{340}},
  \bibinfo{pages}{1545--1548} (\bibinfo{year}{2013}).

\bibitem{Sakaguchi2012SpaceFiber}
\bibinfo{author}{Sakaguchi, J.} \emph{et~al.}
\newblock {Space division multiplexed transmission of 109-Tb/s data signals
  using homogeneous seven-core fiber}.
\newblock \emph{\bibinfo{journal}{Journal of Lightwave Technology}}
  \textbf{\bibinfo{volume}{30}}, \bibinfo{pages}{658--665}
  (\bibinfo{year}{2012}).

\bibitem{Wang2012TerabitMultiplexing}
\bibinfo{author}{Wang, J.} \emph{et~al.}
\newblock {Terabit free-space data transmission employing orbital angular
  momentum multiplexing}.
\newblock \emph{\bibinfo{journal}{Nature Photonics 2012 6:7}}
  \textbf{\bibinfo{volume}{6}}, \bibinfo{pages}{488--496}
  (\bibinfo{year}{2012}).

\bibitem{willner2015optical}
\bibinfo{author}{Willner, A.~E.} \emph{et~al.}
\newblock Optical communications using orbital angular momentum beams.
\newblock \emph{\bibinfo{journal}{Advances in optics and photonics}}
  \textbf{\bibinfo{volume}{7}}, \bibinfo{pages}{66--106}
  (\bibinfo{year}{2015}).

\bibitem{Molina-Terriza2007TwistedPhotons}
\bibinfo{author}{Molina-Terriza, G.}, \bibinfo{author}{Torres, J.~P.} \&
  \bibinfo{author}{Torner, L.}
\newblock {Twisted photons}.
\newblock \emph{\bibinfo{journal}{Nature Physics 2007 3:5}}
  \textbf{\bibinfo{volume}{3}}, \bibinfo{pages}{305--310}
  (\bibinfo{year}{2007}).

\bibitem{erhard2018twisted}
\bibinfo{author}{Erhard, M.}, \bibinfo{author}{Fickler, R.},
  \bibinfo{author}{Krenn, M.} \& \bibinfo{author}{Zeilinger, A.}
\newblock Twisted photons: new quantum perspectives in high dimensions.
\newblock \emph{\bibinfo{journal}{Light: Science \& Applications}}
  \textbf{\bibinfo{volume}{7}}, \bibinfo{pages}{17146--17146}
  (\bibinfo{year}{2018}).

\bibitem{Clark2019InteractingPolaritons}
\bibinfo{author}{Clark, L.~W.} \emph{et~al.}
\newblock {Interacting Floquet polaritons}.
\newblock \emph{\bibinfo{journal}{Nature 2019 571:7766}}
  \textbf{\bibinfo{volume}{571}}, \bibinfo{pages}{532--536}
  (\bibinfo{year}{2019}).

\bibitem{Ivanov2018AdiabaticPolaritons}
\bibinfo{author}{Ivanov, P.~A.}, \bibinfo{author}{Letscher, F.},
  \bibinfo{author}{Simon, J.} \& \bibinfo{author}{Fleischhauer, M.}
\newblock {Adiabatic flux insertion and growing of Laughlin states of cavity
  Rydberg polaritons}.
\newblock \emph{\bibinfo{journal}{Physical Review A}}
  \textbf{\bibinfo{volume}{98}}, \bibinfo{pages}{013847}
  (\bibinfo{year}{2018}).

\bibitem{kimble2008quantum}
\bibinfo{author}{Kimble, H.~J.}
\newblock The quantum internet.
\newblock \emph{\bibinfo{journal}{Nature}} \textbf{\bibinfo{volume}{453}},
  \bibinfo{pages}{1023--1030} (\bibinfo{year}{2008}).

\bibitem{baharudin2013review}
\bibinfo{author}{Baharudin, N.}, \bibinfo{author}{Alsaqour, R.},
  \bibinfo{author}{Shaker, H.}, \bibinfo{author}{Alsaqour, O.} \&
  \bibinfo{author}{Alahdal, T.}
\newblock Review on multiplexing techniques in bandwidth utilization.
\newblock \emph{\bibinfo{journal}{Middle-East Journal of Scientific Research}}
  \textbf{\bibinfo{volume}{18}}, \bibinfo{pages}{1510--1516}
  (\bibinfo{year}{2013}).

\bibitem{sangdeh2019overview}
\bibinfo{author}{Sangdeh, P.~K.} \& \bibinfo{author}{Zeng, H.}
\newblock Overview of Multiplexing Techniques in Wireless Networks.
\newblock In \emph{\bibinfo{booktitle}{Multiplexing}}
  (\bibinfo{publisher}{IntechOpen London}, \bibinfo{year}{2019}).

\bibitem{ishio1984review}
\bibinfo{author}{Ishio, H.}, \bibinfo{author}{Minowa, J.} \&
  \bibinfo{author}{Nosu, K.}
\newblock Review and status of wavelength-division-multiplexing technology and
  its application.
\newblock \emph{\bibinfo{journal}{Journal of lightwave technology}}
  \textbf{\bibinfo{volume}{2}}, \bibinfo{pages}{448--463}
  (\bibinfo{year}{1984}).

\bibitem{ralph2007efficient}
\bibinfo{author}{Ralph, T.}, \bibinfo{author}{Resch, K.} \&
  \bibinfo{author}{Gilchrist, A.}
\newblock Efficient Toffoli gates using qudits.
\newblock \emph{\bibinfo{journal}{Physical Review A}}
  \textbf{\bibinfo{volume}{75}}, \bibinfo{pages}{022313}
  (\bibinfo{year}{2007}).

\bibitem{chen2017realization}
\bibinfo{author}{Chen, D.-X.} \emph{et~al.}
\newblock Realization of quantum permutation algorithm in high dimensional
  Hilbert space.
\newblock \emph{\bibinfo{journal}{Chinese Physics B}}
  \textbf{\bibinfo{volume}{26}}, \bibinfo{pages}{060305}
  (\bibinfo{year}{2017}).

\bibitem{Goyal2013TeleportingScissors}
\bibinfo{author}{Goyal, S.~K.} \& \bibinfo{author}{Konrad, T.}
\newblock {Teleporting photonic qudits using multimode quantum scissors}.
\newblock \emph{\bibinfo{journal}{Scientific Reports 2013 3:1}}
  \textbf{\bibinfo{volume}{3}}, \bibinfo{pages}{1--4} (\bibinfo{year}{2013}).

\bibitem{Garcia-Escartin2008QuantumLight}
\bibinfo{author}{Garc{\'{i}}a-Escart{\'{i}}n, J.~C.} \&
  \bibinfo{author}{Chamorro-Posada, P.}
\newblock {Quantum multiplexing with the orbital angular momentum of light}.
\newblock \emph{\bibinfo{journal}{Physical Review A - Atomic, Molecular, and
  Optical Physics}} \textbf{\bibinfo{volume}{78}}, \bibinfo{pages}{062320}
  (\bibinfo{year}{2008}).

\bibitem{cozzolino2019high}
\bibinfo{author}{Cozzolino, D.}, \bibinfo{author}{Da~Lio, B.},
  \bibinfo{author}{Bacco, D.} \& \bibinfo{author}{Oxenl{\o}we, L.~K.}
\newblock High-dimensional quantum communication: benefits, progress, and
  future challenges.
\newblock \emph{\bibinfo{journal}{Advanced Quantum Technologies}}
  \textbf{\bibinfo{volume}{2}}, \bibinfo{pages}{1900038}
  (\bibinfo{year}{2019}).

\bibitem{Mirhosseini2015High-dimensionalLight}
\bibinfo{author}{Mirhosseini, M.} \emph{et~al.}
\newblock {High-dimensional quantum cryptography with twisted light}.
\newblock \emph{\bibinfo{journal}{New Journal of Physics}}
  \textbf{\bibinfo{volume}{17}}, \bibinfo{pages}{033033}
  (\bibinfo{year}{2015}).

\bibitem{Mair2001EntanglementPhotons}
\bibinfo{author}{Mair, A.}, \bibinfo{author}{Vaziri, A.},
  \bibinfo{author}{Weihs, G.} \& \bibinfo{author}{Zeilinger, A.}
\newblock {Entanglement of the orbital angular momentum states of photons}.
\newblock \emph{\bibinfo{journal}{Nature 2001 412:6844}}
  \textbf{\bibinfo{volume}{412}}, \bibinfo{pages}{313--316}
  (\bibinfo{year}{2001}).

\bibitem{Krenn2014GenerationSystem}
\bibinfo{author}{Krenn, M.} \emph{et~al.}
\newblock {Generation and confirmation of a (100 × 100)-dimensional entangled
  quantum system}.
\newblock \emph{\bibinfo{journal}{Proceedings of the National Academy of
  Sciences of the United States of America}} \textbf{\bibinfo{volume}{111}},
  \bibinfo{pages}{6243--6247} (\bibinfo{year}{2014}).

\bibitem{ren2015free}
\bibinfo{author}{Ren, Y.} \emph{et~al.}
\newblock Free-space optical communications using orbital-angular-momentum
  multiplexing combined with MIMO-based spatial multiplexing.
\newblock \emph{\bibinfo{journal}{Optics letters}}
  \textbf{\bibinfo{volume}{40}}, \bibinfo{pages}{4210--4213}
  (\bibinfo{year}{2015}).

\bibitem{tariq2021orbital}
\bibinfo{author}{Tariq, U.}, \bibinfo{author}{Shahoei, H.},
  \bibinfo{author}{Yang, G.} \& \bibinfo{author}{MacFarlane, D.~L.}
\newblock Orbital Angular Momentum Orthogonality Based Crosstalk Reduction.
\newblock \emph{\bibinfo{journal}{Progress In Electromagnetics Research
  Letters}} \textbf{\bibinfo{volume}{98}}, \bibinfo{pages}{17--25}
  (\bibinfo{year}{2021}).

\bibitem{Shen2022ModeBeams}
\bibinfo{author}{Shen, D.}, \bibinfo{author}{He, T.}, \bibinfo{author}{Yu, X.}
  \& \bibinfo{author}{Zhao, D.}
\newblock {Mode Conversion and Transfer of Orbital Angular Momentum between
  Hermite-Gaussian and Laguerre-Gaussian Beams}.
\newblock \emph{\bibinfo{journal}{IEEE Photonics Journal}}
  \textbf{\bibinfo{volume}{14}} (\bibinfo{year}{2022}).

\bibitem{Liang2019ControllablePhase}
\bibinfo{author}{Liang, G.} \& \bibinfo{author}{Wang, Q.}
\newblock {Controllable conversion between Hermite Gaussian and Laguerre
  Gaussian modes due to cross phase}.
\newblock \emph{\bibinfo{journal}{Optics Express}}
  \textbf{\bibinfo{volume}{27}}, \bibinfo{pages}{10684} (\bibinfo{year}{2019}).

\bibitem{fontaine2019laguerre}
\bibinfo{author}{Fontaine, N.~K.} \emph{et~al.}
\newblock Laguerre-Gaussian mode sorter.
\newblock \emph{\bibinfo{journal}{Nature communications}}
  \textbf{\bibinfo{volume}{10}}, \bibinfo{pages}{1--7} (\bibinfo{year}{2019}).

\bibitem{zhou2018hermite}
\bibinfo{author}{Zhou, Y.} \emph{et~al.}
\newblock Hermite--Gaussian mode sorter.
\newblock \emph{\bibinfo{journal}{Optics letters}}
  \textbf{\bibinfo{volume}{43}}, \bibinfo{pages}{5263--5266}
  (\bibinfo{year}{2018}).

\bibitem{beijersbergen1993astigmatic}
\bibinfo{author}{Beijersbergen, M.~W.}, \bibinfo{author}{Allen, L.},
  \bibinfo{author}{Van~der Veen, H.} \& \bibinfo{author}{Woerdman, J.}
\newblock Astigmatic laser mode converters and transfer of orbital angular
  momentum.
\newblock \emph{\bibinfo{journal}{Optics Communications}}
  \textbf{\bibinfo{volume}{96}}, \bibinfo{pages}{123--132}
  (\bibinfo{year}{1993}).

\bibitem{yao2011orbital}
\bibinfo{author}{Yao, A.~M.} \& \bibinfo{author}{Padgett, M.~J.}
\newblock Orbital angular momentum: origins, behavior and applications.
\newblock \emph{\bibinfo{journal}{Advances in optics and photonics}}
  \textbf{\bibinfo{volume}{3}}, \bibinfo{pages}{161--204}
  (\bibinfo{year}{2011}).

\bibitem{Danaci2016All-opticalMixing}
\bibinfo{author}{Danaci, O.}, \bibinfo{author}{Rios, C.} \&
  \bibinfo{author}{Glasser, R.~T.}
\newblock {All-optical mode conversion via spatially multimode four-wave
  mixing}.
\newblock \emph{\bibinfo{journal}{New Journal of Physics}}
  \textbf{\bibinfo{volume}{18}}, \bibinfo{pages}{073032}
  (\bibinfo{year}{2016}).

\bibitem{Nie2016MultichannelCommunication}
\bibinfo{author}{Nie, S.}, \bibinfo{author}{Yu, S.}, \bibinfo{author}{Cai, S.},
  \bibinfo{author}{Lan, M.} \& \bibinfo{author}{Gu, W.}
\newblock {Multichannel mode conversion and multiplexing based on a single
  spatial light modulator for optical communication}.
\newblock \emph{\bibinfo{journal}{https://doi.org/10.1117/1.OE.55.7.076108}}
  \textbf{\bibinfo{volume}{55}}, \bibinfo{pages}{076108}
  (\bibinfo{year}{2016}).

\bibitem{Shen2022OAMChip}
\bibinfo{author}{Shen, W.~G.}, \bibinfo{author}{Chen, Y.},
  \bibinfo{author}{Wang, H.~M.} \& \bibinfo{author}{Jin, X.~M.}
\newblock {OAM mode conversion in a photonic chip}.
\newblock \emph{\bibinfo{journal}{Optics Communications}}
  \textbf{\bibinfo{volume}{507}}, \bibinfo{pages}{127615}
  (\bibinfo{year}{2022}).

\bibitem{Pires2019OpticalMixing}
\bibinfo{author}{Pires, D.~G.}, \bibinfo{author}{Rocha, J.~C.},
  \bibinfo{author}{Jesus-Silva, A.~J.} \& \bibinfo{author}{Fonseca, E.~J.}
\newblock {Optical mode conversion through nonlinear two-wave mixing}.
\newblock \emph{\bibinfo{journal}{Physical Review A}}
  \textbf{\bibinfo{volume}{100}}, \bibinfo{pages}{043819}
  (\bibinfo{year}{2019}).

\bibitem{cumming1957serrodyne}
\bibinfo{author}{Cumming, R.~C.}
\newblock The serrodyne frequency translator.
\newblock \emph{\bibinfo{journal}{Proceedings of the IRE}}
  \textbf{\bibinfo{volume}{45}}, \bibinfo{pages}{175--186}
  (\bibinfo{year}{1957}).

\bibitem{Piccirillo2009LightCharge}
\bibinfo{author}{Piccirillo, B.}, \bibinfo{author}{Karimi, E.},
  \bibinfo{author}{Santamato, E.} \& \bibinfo{author}{Marrucci, L.}
\newblock {Light propagation in a birefringent plate with topological charge}.
\newblock \emph{\bibinfo{journal}{Optics Letters, Vol. 34, Issue 8, pp.
  1225-1227}} \textbf{\bibinfo{volume}{34}}, \bibinfo{pages}{1225--1227}
  (\bibinfo{year}{2009}).

\bibitem{Piccirillo2011EfficientLinks}
\bibinfo{author}{Piccirillo, B.}, \bibinfo{author}{Karimi, E.},
  \bibinfo{author}{Santamato, E.}, \bibinfo{author}{Marrucci, L.} \&
  \bibinfo{author}{Slussarenko, S.}
\newblock {Efficient generation and control of different-order orbital angular
  momentum states for communication links}.
\newblock \emph{\bibinfo{journal}{JOSA A, Vol. 28, Issue 1, pp. 61-65}}
  \textbf{\bibinfo{volume}{28}}, \bibinfo{pages}{61--65}
  (\bibinfo{year}{2011}).

\bibitem{Karimi2009EfficientQ-plates}
\bibinfo{author}{Karimi, E.}, \bibinfo{author}{Piccirillo, B.},
  \bibinfo{author}{Nagali, E.}, \bibinfo{author}{Marrucci, L.} \&
  \bibinfo{author}{Santamato, E.}
\newblock {Efficient generation and sorting of orbital angular momentum
  eigenmodes of light by thermally tuned q-plates}.
\newblock \emph{\bibinfo{journal}{Applied Physics Letters}}
  \textbf{\bibinfo{volume}{94}}, \bibinfo{pages}{231124}
  (\bibinfo{year}{2009}).

\bibitem{slussarenko2013liquid}
\bibinfo{author}{Slussarenko, S.}, \bibinfo{author}{Piccirillo, B.},
  \bibinfo{author}{Chigrinov, V.}, \bibinfo{author}{Marrucci, L.} \&
  \bibinfo{author}{Santamato, E.}
\newblock Liquid crystal spatial-mode converters for the orbital angular
  momentum of light.
\newblock \emph{\bibinfo{journal}{Journal of Optics}}
  \textbf{\bibinfo{volume}{15}}, \bibinfo{pages}{025406}
  (\bibinfo{year}{2013}).

\bibitem{karimi2009light}
\bibinfo{author}{Karimi, E.}, \bibinfo{author}{Piccirillo, B.},
  \bibinfo{author}{Marrucci, L.} \& \bibinfo{author}{Santamato, E.}
\newblock Light propagation in a birefringent plate with topological charge.
\newblock \emph{\bibinfo{journal}{Optics letters}}
  \textbf{\bibinfo{volume}{34}}, \bibinfo{pages}{1225--1227}
  (\bibinfo{year}{2009}).

\bibitem{wei2019generating}
\bibinfo{author}{Wei, D.} \emph{et~al.}
\newblock Generating controllable Laguerre-Gaussian laser modes through
  intracavity spin-orbital angular momentum conversion of light.
\newblock \emph{\bibinfo{journal}{Physical Review Applied}}
  \textbf{\bibinfo{volume}{11}}, \bibinfo{pages}{014038}
  (\bibinfo{year}{2019}).

\bibitem{Nakagawa2020LaguerreGaussianMicrocavity}
\bibinfo{author}{Nakagawa, K.}, \bibinfo{author}{Yamane, K.},
  \bibinfo{author}{Morita, R.} \& \bibinfo{author}{Toda, Y.}
\newblock {Laguerre–Gaussian vortex mode generation from astigmatic
  semiconductor microcavity}.
\newblock \emph{\bibinfo{journal}{Applied Physics Express}}
  \textbf{\bibinfo{volume}{13}}, \bibinfo{pages}{042001}
  (\bibinfo{year}{2020}).

\bibitem{stone2021optical}
\bibinfo{author}{Stone, M.}, \bibinfo{author}{Suleymanzade, A.},
  \bibinfo{author}{Taneja, L.}, \bibinfo{author}{Schuster, D.~I.} \&
  \bibinfo{author}{Simon, J.}
\newblock Optical mode conversion in coupled Fabry--Perot resonators.
\newblock \emph{\bibinfo{journal}{Optics Letters}}
  \textbf{\bibinfo{volume}{46}}, \bibinfo{pages}{21--24}
  (\bibinfo{year}{2021}).

\bibitem{lu2013nanophotonic}
\bibinfo{author}{Lu, J.} \& \bibinfo{author}{Vu{\v{c}}kovi{\'c}, J.}
\newblock Nanophotonic computational design.
\newblock \emph{\bibinfo{journal}{Optics express}}
  \textbf{\bibinfo{volume}{21}}, \bibinfo{pages}{13351--13367}
  (\bibinfo{year}{2013}).

\bibitem{Li2018MultimodePhotonics}
\bibinfo{author}{Li, C.}, \bibinfo{author}{Liu, D.} \& \bibinfo{author}{Dai,
  D.}
\newblock {Multimode silicon photonics}.
\newblock \emph{\bibinfo{journal}{Nanophotonics}} \textbf{\bibinfo{volume}{8}},
  \bibinfo{pages}{227--247} (\bibinfo{year}{2018}).

\bibitem{Wang2019CompactStructure}
\bibinfo{author}{Wang, H.} \emph{et~al.}
\newblock {Compact Silicon Waveguide Mode Converter Employing Dielectric
  Metasurface Structure}.
\newblock \emph{\bibinfo{journal}{Advanced Optical Materials}}
  \textbf{\bibinfo{volume}{7}}, \bibinfo{pages}{1801191}
  (\bibinfo{year}{2019}).

\bibitem{Karabchevsky2019SpatialWaveguides}
\bibinfo{author}{Karabchevsky, A.} \& \bibinfo{author}{Greenberg, Y.}
\newblock {Spatial eigenmodes conversion with metasurfaces engraved in silicon
  ridge waveguides}.
\newblock \emph{\bibinfo{journal}{Applied Optics, Vol. 58, Issue 22, pp.
  F21-F25}} \textbf{\bibinfo{volume}{58}}, \bibinfo{pages}{F21--F25}
  (\bibinfo{year}{2019}).

\bibitem{Wajih2019ASlots}
\bibinfo{author}{Wajih, M.} \emph{et~al.}
\newblock {A compact silicon-based TM0-to-TM2 mode-order converter using
  shallowly-etched slots}.
\newblock \emph{\bibinfo{journal}{Journal of Optics}}
  \textbf{\bibinfo{volume}{22}}, \bibinfo{pages}{015802}
  (\bibinfo{year}{2019}).

\bibitem{Wang2020Ultra-compactSlots}
\bibinfo{author}{Wang, H.} \emph{et~al.}
\newblock {Ultra-compact silicon mode-order converters based on dielectric
  slots}.
\newblock \emph{\bibinfo{journal}{Optics Letters, Vol. 45, Issue 13, pp.
  3797-3800}} \textbf{\bibinfo{volume}{45}}, \bibinfo{pages}{3797--3800}
  (\bibinfo{year}{2020}).

\bibitem{Zhang2020On-chipCrosstalk}
\bibinfo{author}{Zhang, B.} \emph{et~al.}
\newblock {On-chip silicon shallowly etched TM0-to-TM1 mode-order converter
  with high conversion efficiency and low modal crosstalk}.
\newblock \emph{\bibinfo{journal}{JOSA B, Vol. 37, Issue 5, pp. 1290-1297}}
  \textbf{\bibinfo{volume}{37}}, \bibinfo{pages}{1290--1297}
  (\bibinfo{year}{2020}).

\bibitem{Zhu2021Silicon-BasedMetasurface}
\bibinfo{author}{Zhu, C.} \emph{et~al.}
\newblock {Silicon-Based TM0-to-TM3 Mode-Order Converter Using On-Chip
  Shallowly Etched Slot Metasurface}.
\newblock \emph{\bibinfo{journal}{Photonics 2021, Vol. 8, Page 95}}
  \textbf{\bibinfo{volume}{8}}, \bibinfo{pages}{95} (\bibinfo{year}{2021}).

\bibitem{Miller2012Ultra-compactEffect}
\bibinfo{author}{Miller, D. A.~B.}, \bibinfo{author}{Fan, S.} \&
  \bibinfo{author}{Liu, V.}
\newblock {Ultra-compact photonic crystal waveguide spatial mode converter and
  its connection to the optical diode effect}.
\newblock \emph{\bibinfo{journal}{Optics Express, Vol. 20, Issue 27, pp.
  28388-28397}} \textbf{\bibinfo{volume}{20}}, \bibinfo{pages}{28388--28397}
  (\bibinfo{year}{2012}).

\bibitem{Huang2006AnConverter}
\bibinfo{author}{Huang, Y.}, \bibinfo{author}{Xu, G.} \& \bibinfo{author}{Ho,
  S.~T.}
\newblock {An ultracompact optical mode order converter}.
\newblock \emph{\bibinfo{journal}{IEEE Photonics Technology Letters}}
  \textbf{\bibinfo{volume}{18}}, \bibinfo{pages}{2281--2283}
  (\bibinfo{year}{2006}).

\bibitem{Dai2012ModeWaveguides}
\bibinfo{author}{Dai, D.} \emph{et~al.}
\newblock {Mode conversion in tapered submicron silicon ridge optical
  waveguides}.
\newblock \emph{\bibinfo{journal}{Optics Express, Vol. 20, Issue 12, pp.
  13425-13439}} \textbf{\bibinfo{volume}{20}}, \bibinfo{pages}{13425--13439}
  (\bibinfo{year}{2012}).

\bibitem{dai2013silicon}
\bibinfo{author}{Dai, D.}, \bibinfo{author}{Wang, J.} \& \bibinfo{author}{He,
  S.}
\newblock Silicon multimode photonic integrated devices for on-chip
  mode-division-multiplexed optical interconnects (invited review).
\newblock \emph{\bibinfo{journal}{Progress In Electromagnetics Research}}
  \textbf{\bibinfo{volume}{143}}, \bibinfo{pages}{773--819}
  (\bibinfo{year}{2013}).

\bibitem{Frandsen2014TopologyMaterial}
\bibinfo{author}{Frandsen, L.~H.} \emph{et~al.}
\newblock {Topology optimized mode conversion in a photonic crystal waveguide
  fabricated in silicon-on-insulator material}.
\newblock \emph{\bibinfo{journal}{Optics Express, Vol. 22, Issue 7, pp.
  8525-8532}} \textbf{\bibinfo{volume}{22}}, \bibinfo{pages}{8525--8532}
  (\bibinfo{year}{2014}).

\bibitem{Chen2005WaveguideCrystals}
\bibinfo{author}{Chen, G.} \& \bibinfo{author}{Kang, J.~U.}
\newblock {Waveguide mode converter based on two-dimensional photonic
  crystals}.
\newblock \emph{\bibinfo{journal}{Optics Letters, Vol. 30, Issue 13, pp.
  1656-1658}} \textbf{\bibinfo{volume}{30}}, \bibinfo{pages}{1656--1658}
  (\bibinfo{year}{2005}).

\bibitem{Schine2016SyntheticPhotons}
\bibinfo{author}{Schine, N.}, \bibinfo{author}{Ryou, A.},
  \bibinfo{author}{Gromov, A.}, \bibinfo{author}{Sommer, A.} \&
  \bibinfo{author}{Simon, J.}
\newblock {Synthetic Landau levels for photons}.
\newblock \emph{\bibinfo{journal}{Nature}} \textbf{\bibinfo{volume}{534}},
  \bibinfo{pages}{671--675} (\bibinfo{year}{2016}).

\bibitem{tanji2011interaction}
\bibinfo{author}{Tanji-Suzuki, H.} \emph{et~al.}
\newblock Interaction between atomic ensembles and optical resonators:
  Classical description.
\newblock In \emph{\bibinfo{booktitle}{Advances in atomic, molecular, and
  optical physics}}, vol.~\bibinfo{volume}{60}, \bibinfo{pages}{201--237}
  (\bibinfo{publisher}{Elsevier}, \bibinfo{year}{2011}).

\bibitem{Clark2020ObservationLight}
\bibinfo{author}{Clark, L.~W.}, \bibinfo{author}{Schine, N.},
  \bibinfo{author}{Baum, C.}, \bibinfo{author}{Jia, N.} \&
  \bibinfo{author}{Simon, J.}
\newblock {Observation of Laughlin states made of light}.
\newblock \emph{\bibinfo{journal}{Nature 2020 582:7810}}
  \textbf{\bibinfo{volume}{582}}, \bibinfo{pages}{41--45}
  (\bibinfo{year}{2020}).

\bibitem{jaffe2021aberrated}
\bibinfo{author}{Jaffe, M.}, \bibinfo{author}{Palm, L.}, \bibinfo{author}{Baum,
  C.}, \bibinfo{author}{Taneja, L.} \& \bibinfo{author}{Simon, J.}
\newblock Aberrated optical cavities.
\newblock \emph{\bibinfo{journal}{Physical Review A}}
  \textbf{\bibinfo{volume}{104}}, \bibinfo{pages}{013524}
  (\bibinfo{year}{2021}).

\end{thebibliography}

	\renewcommand{\appendixname}{Methods}
	
	\setcounter{equation}{0}
	\setcounter{figure}{0}
	\renewcommand{\theequation}{M\arabic{equation}}
	\renewcommand{\thefigure}{M\arabic{figure}}
	
	\clearpage
	
	\setcounter{secnumdepth}{2}
	
	\section*{Methods}
	
	\noindent\textbf{Experimental setup.} The main ingredients used in this work are the twisted cavity, atomic sample, $780$~nm probe beam, and $1529$~nm modulation beam. Probe photons were converted between eigenmodes of the twisted cavity via passage through the sample of atoms, whose energy levels were spatiotemporally modulated to create a spatiotemporally-varying optical susceptibilty akin to sculpting a phase plate out of the atomic sample.
	
	The twisted cavity used in this work is the same as that described in~\cite{Clark2020ObservationLight} and~\cite{Clark2019InteractingPolaritons}. The eigenmodes of this cavity are non-degenerate Laguerre-Gaussian (LG) modes at the lower cavity waist which coincides with the position of the atomic sample. In reality, paraxial astigmatism distort the LG modes at other positions along the cavity axis~\cite{jaffe2021aberrated}. Thus, in order to incouple to the $l=3$ eigenmode at the location of the atoms, $780$~nm probe light is injected in the Ince-Gaussian spatial mode profile depicted in Figure~\ref{fig:ExpSetup}. The transverse mode spacing between every third orbital angular momentum mode ($l=0,3,6...$) is about 65 MHz with slight variation depending on the choice of free spectral range. The free spectral range is 2.5 GHz. The four cavity mirrors were coated and supplied by LAYERTEC GmbH. As these mirrors are sufficiently reflective at both $780$~nm and $1529$~nm, cavity modes exist for both the probe and modulation beams where further specifications are listed in Table~\ref{table:specs}. The lower waist size at $1529$~nm is related to the lower waist size at $780$~nm by a factor of $\sqrt{1529/780}$.
	
	This work requires probe light to be coupled into the $l=3$ cavity mode and modulation light to be coupled into both the $l=0$ and $l=3$ cavity modes. Due to the aforementioned astigmatism, we inject probe photons with an Ince-Gaussian spatial profile corresponding with $l=3$ at the lower cavity waist. This profile is acquired by using a digital micromirror device (DMD) to shape a preliminary probe beam as depicted in Figure~\ref{fig:ExpSetup}. We use an electro-optic modulator (EOM) to inject a frequency-modulated $l=0$ modulation beam off-center from the cavity axis to couple to the $l=0$ and $l=3$ cavity modes. While the off-center injection of an $l=0$ transverse mode has spatial overlap with many other transverse modes, we couple only to the $l=0$ and $l=3$ cavity modes by frequency discrimination. Off-center injection is not particularly efficient, but this inefficiency was compensated by the large amount of $1529$~nm power we had at our disposal ($>1$ W).
	
	In order to have simultaneous injection of the $l=3$, $780$~nm mode and $l=0$ and $l=3$, $1529$~nm modes, we tune the lockpoint frequency of the $1529$~nm laser and the cavity length using a piezoelectric actuator. We first tune the cavity length to transmit the $l=3$, $780$~nm mode, then change the lockpoint frequency of the $1529$~nm laser such that the carrier and one sideband generated by the EOM are resonant with the $l=0$ and $l=3$ modes. The modulation depth of the EOM is controlled by a variable attenuator to sweep the relative $l=0$ to $l=3$ power. For given values of the collective cooperativity and cavity-atom detuning, we optimize $\mathcal{E}_{3\rightarrow0}$ through iterative fine-tuning scans of the $1529$~nm lockpoint frequency, EOM modulation frequency, and EOM modulation depth.
	
	\noindent\textbf{Laser detunings and polarization.} In this work, the $l=3$, $780$~nm probe beam is 130 MHz detuned from the $5S_{1/2}\leftrightarrow5P_{3/2}$ transition (see SI~\ref{SI:Det}) and the $1529$~nm modulation beam is about 14 GHz detuned from the $5P_{3/2}\leftrightarrow4D_{5/2}$ transition. This 14 GHz detuning was selected for several reasons. First, the $5P_{3/2}\leftrightarrow4D_{3/2}$ transition is only 13.4 GHz higher than the $5P_{3/2}\leftrightarrow4D_{5/2}$ transition. We utilize only circular polarization for the probe and modulation beams in this work, isolating only the stretched states of $\left|5S_{1/2}, F=2\right\rangle$, $\left|5P_{3/2}, F=3\right\rangle$, and $\left|4D_{5/2}, F=4\right\rangle$ assuming perfect polarization and optical pumping. In the event of imperfection, the relatively large detuning of 14 GHz + 13.4 GHz from the $5P_{3/2}\leftrightarrow4D_{3/2}$ transition suppresses mixing of the $4D_{3/2}$ state that could potentially complicate the mode conversion process. Second, $\Omega/(2\pi)\ll$14 GHz for all $\Omega$ used in this work. This condition simplifies the intuition and calculations behind the mode conversion process: for $\Omega$ much less than the detuning from the $5P_{3/2}\leftrightarrow4D_{5/2}$ transition, the $4D_{5/2}$ state remains essentially unpopulated. Thus, the modulation beam can be thought to virtually excite the $4D_{5/2}$ state to convert $l=3$ probe photons to $l=0$. In calculations, the $4D_{5/2}$ state can be adiabatically eliminated, reducing the coupling to the $4D_{5/2}$ state to effective couplings in the Hamiltonian (see SI~\ref{SI:Model}). Third, 14 GHz was convenient given our available frequency sources and high power at $1529$~nm.
	
	\noindent\textbf{Atom number.} The peak cavity-atom coupling between a single $^{87}$Rb atom and $l=0$ mode at $780$~nm is the same as that in~\cite{Clark2020ObservationLight}: $g_{\mathrm{single}}=2\pi\times0.58$ MHz. Given this information, it is possible to estimate the number of atoms from the dispersive shift of the twisted cavity transmission feature in spectra measurements for an unmodulated atomic sample. This shift depends on $Ng_{\mathrm{single}}^2$ where N is the atom number. In this work, we estimate an atom number of 500 for measurements with the lowest atom number ($N\eta=70$) and 3500 for measurements with the highest atom number ($N\eta=560$). $N\eta$ is equivalent to $4Ng_{\mathrm{single}}^2/\kappa\Gamma$ where $\kappa=2\pi\times1.6$ MHz (the cavity linewidth at $780$~nm) and $\Gamma=2\pi\times6$ MHz (the linewdith of the $5P_{3/2}$ state).
	
	\noindent\textbf{Additional mode conversion.} While the majority of this work focused exclusively on $l=3\rightarrow0$ conversion, we briefly examined $l=0\rightarrow3$ conversion. For identical parameters that yielded $l=3\rightarrow0$ conversion near $\mathcal{E}_{3\rightarrow0}$=1, we observed $l=0\rightarrow3$ conversion near $\mathcal{E}_{3\rightarrow0}$=0.5. While this behavior has yet to be understood, it may arise from the unequal detunings of the $l=0$ and $l=3$ modes to the $5P_{3/2}$ state. In early exploratory measurements of this work, we also observed polarization conversion between two $l=0$ polarization modes of the twisted cavity under a slightly different atomic modulation scheme. Instead of modulating the atoms with the $l=0$ and $l=3$ $1529$~nm modes separated by the transverse mode splitting frequency, we modulated the atoms with the two $l=0$ $1529$~nm polarization modes separated by the polarization mode splitting frequency. The polarization conversion efficiency was not rigorously quantified, but polarization conversion is mentioned here to demonstrate proof of concept.
	
	\renewcommand{\appendixname}{Extended Data}
	
	\setcounter{equation}{0}
	\setcounter{figure}{0}
	\renewcommand{\theequation}{Extended Data \arabic{equation}}
	\renewcommand{\thefigure}{Extended Data \arabic{figure}}

	\setcounter{secnumdepth}{2}
	
	\begin{figure*}
		\centering
		\includegraphics[width=1.0\textwidth]{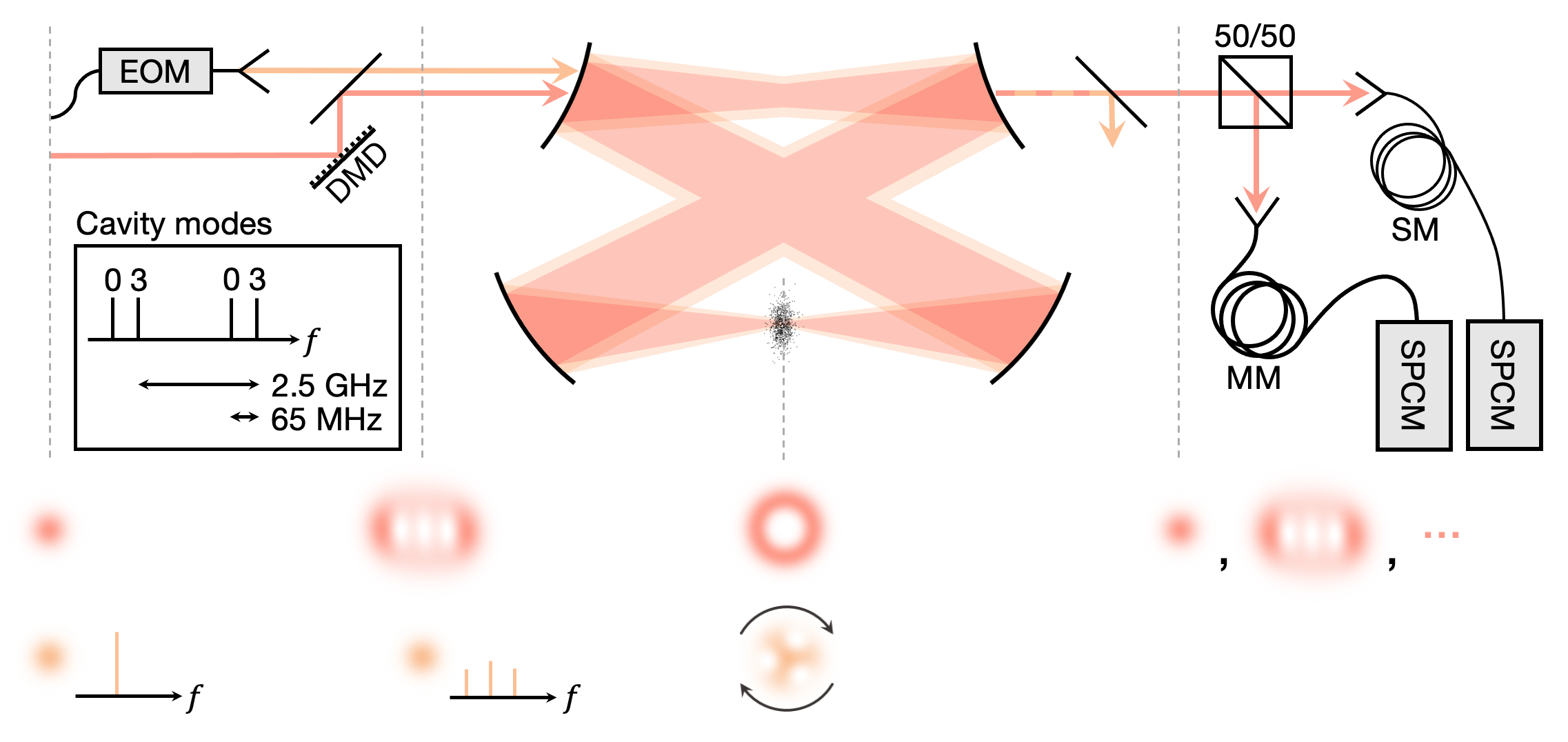}
		\caption{\textbf{Beam preparation and measurement.} From left to right, this figure depicts the preparation of $780$~nm (red) and $1529$~nm light (orange) for injection into the twisted cavity, the $780$~nm and $1529$~nm cavity modes at the lower cavity waist, and the detection of the transmitted $780$~nm light. The $780$~nm light begins as a Gaussian beam which is shaped into an Ince-Gaussian mode by a digital micromirror device (DMD). Due to paraxial astigmatism, this mode evolves into an $l=3$ LG mode at the lower cavity waist due where it couples to the atomic sample. Light eventually leaks out of the cavity and passes through a 50/50 non-polarizing beamsplitter cube, splitting the light between two paths: one through a multi-mode (MM) fiber, and one through a single mode (SM) fiber. Each fiber terminates at a single-photon counting module (SPCM). The SM path is used to detect only $l=0$ cavity photons, whereas the MM path detects all cavity photons. The $1529$~nm light also begins as a Gaussian beam. It passes through an electro-optic modulator (EOM), acquiring frequency sidebands at the frequency difference between the $l=0$ and $l=3$ modes (65 MHz). This light is then injected slightly off-center from the cavity axis which spatially couples the incident $1529$~nm light to a forest of modes, but $l=0$ and $l=3$ are isolated by frequency discrimination. The superposed $l=0$ and $l=3$ modes form a rotating, three-holed profile at the lower cavity waist that spatiotemporally modulates the atomic sample. The $1529$~nm light eventually leaks out of the cavity as well and is filtered out from the detection path.
			\label{fig:ExpSetup}}
	\end{figure*}

	\renewcommand{\tocname}{Supplementary Information}
	\renewcommand{\appendixname}{Supplement}
	
	\setcounter{equation}{0}
	\setcounter{figure}{0}
	\renewcommand{\theequation}{S\arabic{equation}}
	\renewcommand{\thefigure}{S\arabic{figure}}
	
	\incltocpage
	\clearpage
	
	\tableofcontents
	\appendix
	\setcounter{secnumdepth}{2}
	
	\section{Experiment}
	
	\subsection{Definition of $\mathcal{E}_{3\rightarrow0}$}
	\label{SI:Calib}
	
	As illustrated in Figure~\ref{fig:ExpSetup}, the output of the cavity is split into two paths by a 50/50 beamsplitter: one leading to a multimode fiber, and one leading to a single mode fiber. The single mode path collects only $l=0$ light by filtering out higher order modes and the multimode path collects $l=0$, $l=3$, and any other modes which may be present (see SI~\ref{SI:actual0} for why we do not see other modes). The ends of each fiber connect to separate single photon counting modules (SPCMs). Data for each SPCM is collected simultaneously, after which scale factors are applied in post-processing to account for the nonlinearity of the SPCMs and count rate imbalance due to mismatched fiber incoupling efficiencies. To acquire $\mathcal{E}_{3\rightarrow0}$, the internal conversion efficiency from $l=3$ to $l=0$, the post-processed $l=0$ count rate is normalized to the post-processed $l=3$ bare cavity count rate then scaled up by a factor of 4. This factor of 4 arises from the double-ended nature of our cavity, meaning light can leak out one of two mirrors of the cavity (see SI~\ref{SI:ImMatch}). In reality, the cavity is comprised of four mirrors, but two of the mirrors are high reflectors at $780$~nm and so we do not consider these as significant leakage ports. 
	
	In a two-mirror cavity whose mirrors are lossless and equal reflectance, the $l=3$ bare cavity output power is equivalent to the input power assuming perfect spatial incoupling to the cavity. However, our cavity mirrors induce loss as a result of scattering, absorption, and imperfections on the mirror surface such as dust. An estimate of the loss can be derived from the measured finesse and mirror reflectance. The finesse of a two-mirror cavity comprised of identical mirrors with low-loss $A$ and transmissivity $T$ is $2\pi/(2A+2T)$. Given the finesse and reflectance specifications at 780 nm as listed in Table~\ref{table:specs}, we expect the loss per mirror to be about 750 ppm, which corresponds to a maximum $l=3$ bare cavity output power of $(1+A/T)^{-2}=30\%$ of the input power. Thus, the external, or end-to-end, efficiency for $l=3$ to $l=0$ conversion is realistically $\mathcal{E}_{3\rightarrow0}\times\frac{1}{4}\times30\%=7.5\%$ at maximum. This calculation ignores imperfect cavity incoupling which can be corrected for externally with mode-matching optics. However, $\frac{1}{4}\rightarrow1$ in single-ended cavities and $30\%\rightarrow\sim100\%$ for low loss mirrors, leaving significant room to increase the external conversion efficiency to near-100\% in hypothetical future variants of the method presented in this paper.
	
	\subsection{Definition of $\Omega$}
	\label{SI:DefOmega}
	
	The $1529$~nm modulation beam is comprised of an $l=0$ component and an $l=3$ component. In this work, we use $\Omega$ to denote the Rabi frequency of the $l=3$ component which has a direct proportionality to the $l=0$ Rabi frequency. The numerical value of $\Omega$ is estimated through measurement of the cavity line shift in the dispersive regime due to the AC Stark shift provided by a $1529$~nm $l=0$ only. Scale factors are applied to account for differences in the $l=0$ and $l=3$ cavity incoupling efficiencies, spectral redistribution given by the EOM depicted in Figure~\ref{fig:ExpSetup}, and nonlinearity of the acousto-optic modulator used to control the modulation beam intensity. We estimate the $l=3$ Rabi frequency is about 1.7 times higher than the $l=0$ Rabi frequency. At maximum $\Omega/(2\pi)$ of 3.5 GHz, we estimate the total incoupled power is on the order of 1 mW which is then cavity enhanced by a factor of $\mathcal{F}_{1529}/\pi=420$ where $\mathcal{F}_{1529}$ is the cavity finesse at $1529$~nm.
	
	\subsection{Confirmation of an $l=0$ converted output}
	\label{SI:actual0}
	
\begin{figure}
		\centering
		\includegraphics[width=1.0\linewidth]{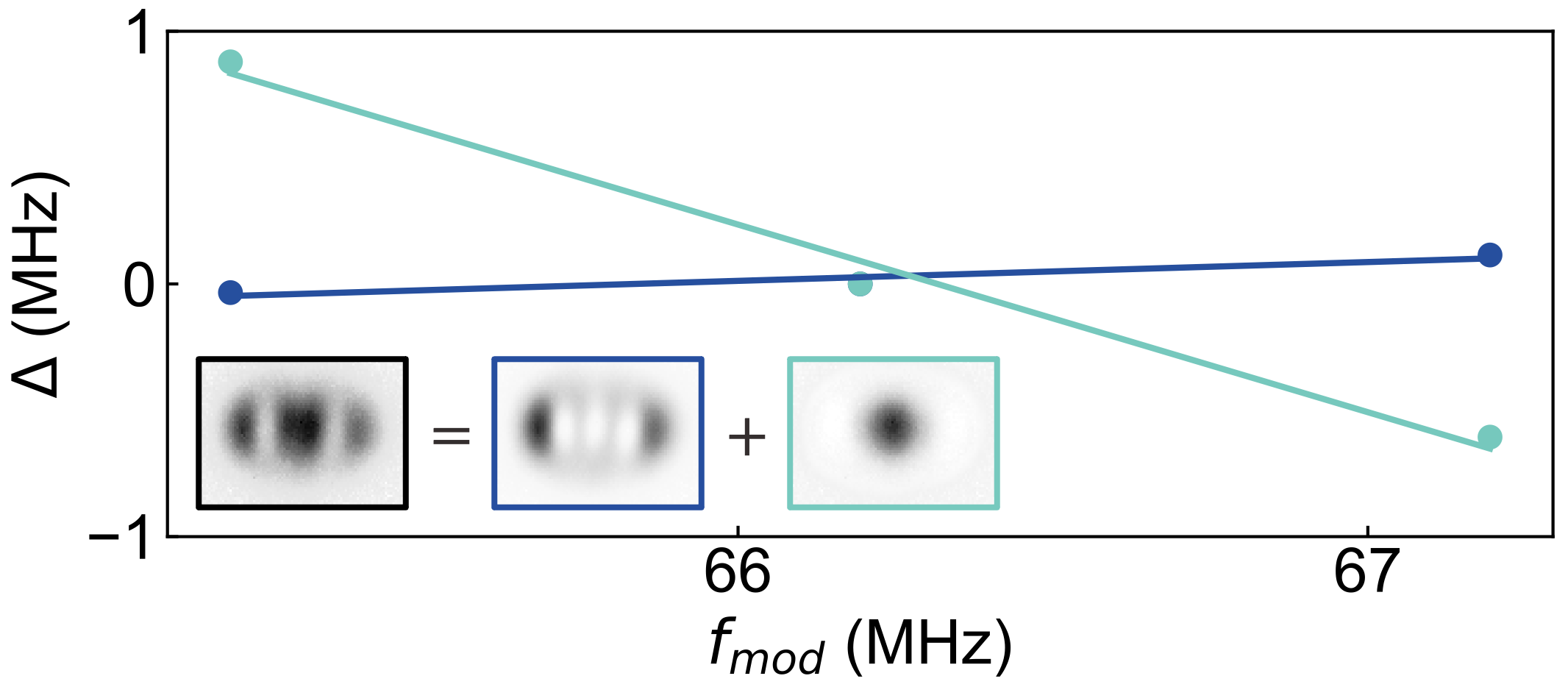}
		\caption{\textbf{Spatial and frequency analysis of the cavity output.} Imaging the total cavity output (black) reveals both unconverted $l=3$ light (blue) and converted $l=0$ light (teal) with conversion efficiency numbers comparable to that of our spectra data. Here, $\Delta$ is the frequency difference between the output $l=3(0)$ mode and the bare cavity $l=3(0)$ mode. Both $l=3$ and $l=0$ light appear at their bare cavity mode frequencies for an atomic modulation frequency, $f_{mod}$, equal to the transverse mode splitting. If $f_{mod}$ is varied slightly, we observe the $l=0$ frequency vary similarly while the $l=3$ frequency remains largely unchanged. Few values of $f_{mod}$ were considered here as this data was collected primarily to confirm the frequencies of the $l=3$ and $l=0$ outputs, and variation of $f_{mod}$ past the scale of the cavity linewidth results in very little conversion as light barely enters the cavity.
			\label{fig:FreqIm}}
	\end{figure}

	Figures~\ref{fig:ConvSpectra} and~\ref{fig:ConvSpatialFreq} depict measurements of converted $l=0$ photons which we verify via imaging and frequency measurements. If light is indeed converted to the $l=0$ eigenmode of the cavity, then we should detect photons with a Gaussian spatial profile and frequency equivalent to that of the bare cavity $l=0$ eigenmode. The lower left corner of Figure~\ref{fig:FreqIm} depicts an image of the cavity output which has been averaged over 200 experimental runs and decomposed into its $l=0$ and $l=3$ constituents for the maximum values of $\Omega$ and $N\eta$ used in this work ($\Omega/(2\pi)=3.5$ GHz and $N\eta=560$). Note that these images were captured for a singular, fixed probe frequency and the image for $l=3$ does not appear LG. While this mode is LG at the location of the atomic sample, it emerges Ince-Gaussian due to astigmatism in the cavity (see~\cite{jaffe2021aberrated} and SI of~\cite{Clark2020ObservationLight}). In order to decompose the image of the cavity output into its $l=0$ and $l=3$ constituents, a bare cavity $l=3$ image was captured, scaled, then subtracted from the cavity output image. Sums were calculated for the cavity output and the bare cavity $l=3$ images over the same small patch centered on the leftmost lobe in each image; the bare cavity $l=3$ image was scaled by the ratio of these sums. The subtraction of the scaled bare cavity $l=3$ image reveals a Gaussian profile expected of the $l=0$ mode. Note that higher order cavity eigenmodes, such as $l=6$ and those with radial nodes, are not observed in imaging. As the modulation beam plausibly induces additional couplings to these modes, we suspect they may be suppressed as a result of their higher detuning from the $5S_{1/2}\rightarrow5P_{3/2}$ transition. We observed nonzero $l=3$ output due to impedance matching considerations as detailed in~\ref{SI:ImMatch}, and quantitative comparison in imaging further supports conversion from $l=3$ to $l=0$ near $\mathcal{E}_{3\rightarrow0}=1$.
	
	Figure~\ref{fig:FreqIm} also depicts the dependence of the $l=0$ and $l=3$ output frequencies on the modulation frequency for a singular, fixed probe frequency. To measure the frequencies of the $l=0$ and $l=3$ constituents, the twisted cavity output is sent through a 2-mirror filter cavity whose length is controllably scanned using a piezoelectric actuator and side-of-fringe lock to an additional laser. This 2-mirror cavity acts as a frequency ruler that could spatially discriminate between modes. For varying $1529$~nm modulation frequencies ($f_{mod}$), we measured the frequency differences ($\Delta$) of the converted $l=0$ and unconverted $l=3$ outputs relative to each of their bare twisted cavity frequencies. Not only did we observe the converted $l=0$ and unconverted $l=3$ to be equal to their bare twisted cavity frequencies for $f_{mod}$ equal to the transverse mode splitting (near 66 MHz for this choice of twisted cavity free spectral range), but we observed the influence of the $1529$~nm modulation on the frequency of the converted $l=0$ output. The converted $l=0$ output frequency changed near-linearly with $f_{mod}$ within about one twisted cavity linewidth at $1529$~nm. Measurements were not collected beyond one twisted cavity linewidth, as the conversion efficiency drops significantly here due to insufficient $1529$~nm power entering the cavity resulting in poor modulation of the atomic sample.

	\subsection{Impedance matching}
	\label{SI:ImMatch}
	
	\begin{figure*}
		\centering
		\includegraphics[width=1.0\linewidth]{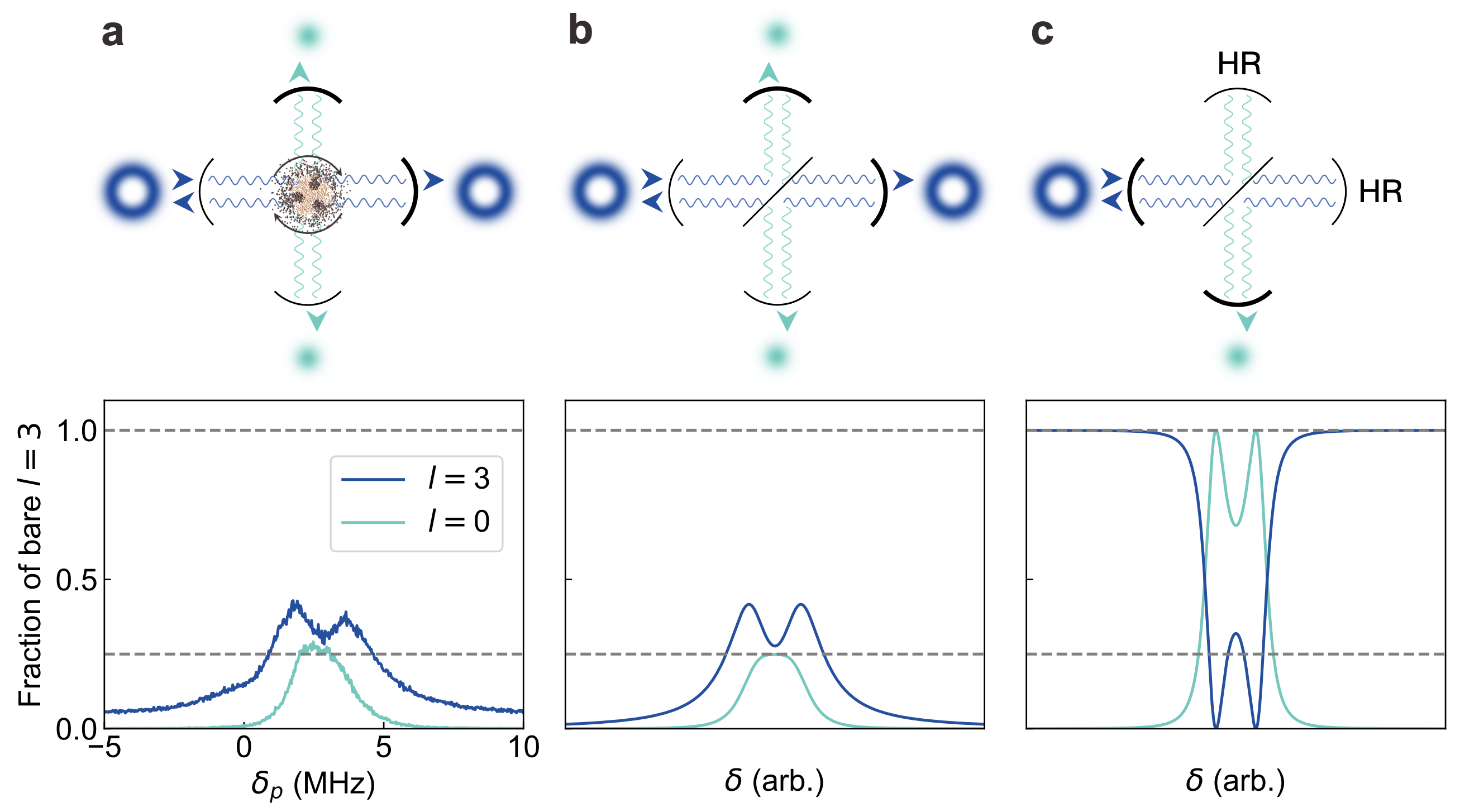}
		\caption{\textbf{Cavity impedance matching and conversion efficiency.} This figure depicts a reinterpreted layout of the twisted cavity and its corresponding transmission spectra. Light can leak out only two of the four cavity mirrors, rendering our cavity double-ended. This work describes two cavity modes that are coupled by a modulated atomic sample, which is modeled here as two crossed, identical cavities that each host one of the spatial modes and are coupled by a coupling element. The cavities need not be visualized as crossed, but they are here for visibility and ease of considering the beam splitter relations of the coupling element. The plots of \textbf{a}, \textbf{b}, and \textbf{c} depict experimental transmissions for the highest values of $\Omega$ and $N\eta$ used in this work, simulated transmissions for a double-ended cavity (akin to the cavity used in this work), and simulated transmissions for a single-ended cavity, respectively. Transmission for $l=3$ was acquired by subtracting the SM signal from the MM signal, including appropriate scale factors to account for coupling efficiency differences and the nonlinearity of the SPCMs (see Fig.~\ref{fig:ExpSetup}). The background level of the $l=3$ transmission is higher than that of $l=0$ transmission as the MM fiber lets in more ambient light. The shapes and numerical values of the transmission curves in \textbf{a} and \textbf{b} are directly comparable. The shapes are a result from the coupling between modes, where coupled modes ‘split' in general, and the numerical values are a result of the double-ended nature of the cavity. If we increase the reflectivity of the coupling element in \textbf{b}, both modes ‘split' further in the spectra, but the $l=0$ transmission never exceeds 25\%. \textbf{c}, If we replace one of the cavity mirrors with a high reflector while all other parameters remain constant, we alter the cavity from double-ended to single-ended and the maximum external efficiency increases from 25\% to 100\%.
			\label{fig:ImpMatch}}
	\end{figure*}
	
	Reference~\cite{stone2021optical} and its appendix are excellent examples of how cavity impedance matching affects the transmission and conversion of cavity modes. Here, we will follow a similar formalism to illustrate why the external conversion efficiency is limited to 25$\%$ in a double-ended cavity and how extension to a single-ended cavity should enable an external conversion efficiency of 100\% for lossless cavity mirrors.

	Figure~\ref{fig:ImpMatch} depicts a reinterpreted layout of the four-mirror twisted cavity. As two of the four mirrors are highly reflective (HR), the twisted cavity is effectively reduced to a two-mirror cavity. As the cavity hosts two coupled eigenmodes in the context of this work, we can model the coupled eigenmodes as two coupled two-mirror cavities where each cavity hosts an eigenmode and the left (right) mirror reflection coefficient is the same as the bottom (top) mirror reflection coefficient $r_1$ ($r_2$). The coupling element is the modulated atomic sample which can be modeled as some partially-reflective optic that obeys the beam splitter relations and has reflection coefficient $r$. Assuming lossless mirrors, $r=\sqrt{1-t^2}$, $r_1=\sqrt{1-t_1^2}$, and $r_2=\sqrt{1-t_2^2}$ where $t$, $t_1$, and $t_2$ are the corresponding transmission coefficients. We now solve the following system of equations to ultimately calculate the $l=3$ and $l=0$ transmissions plotted in Figures~\ref{fig:ImpMatch}b and~\ref{fig:ImpMatch}c:
	
	\begin{align*}
		\mathrm{E}_{1\mathrm{AR}} & = (\mathrm{E}_{1\mathrm{AL}}r_1+\mathrm{E}_{0\mathrm{A}}t_1)e^{i\delta} \\
		\mathrm{E}_{1\mathrm{AL}} & = (\mathrm{E}_{2\mathrm{BL}}r+\mathrm{E}_{2\mathrm{AL}}t)e^{i\delta} \\
		\mathrm{E}_{2\mathrm{AR}} & = (-\mathrm{E}_{1\mathrm{BR}}r+\mathrm{E}_{1\mathrm{AR}}t)e^{i\delta} \\
		\mathrm{E}_{2\mathrm{AL}} & = -\mathrm{E}_{2\mathrm{AR}}r_2e^{i\delta} \\
		\mathrm{E}_{1\mathrm{BR}} & = -\mathrm{E}_{1\mathrm{BL}}r_1e^{i\delta} \\
		\mathrm{E}_{1\mathrm{BL}} & = (-\mathrm{E}_{2\mathrm{AL}}r+\mathrm{E}_{2\mathrm{BL}}t)e^{i\delta} \\
		\mathrm{E}_{2\mathrm{BR}} & = (\mathrm{E}_{1\mathrm{AR}}r+\mathrm{E}_{1\mathrm{BR}}t)e^{i\delta} \\
		\mathrm{E}_{2\mathrm{BL}} & = \mathrm{E}_{2\mathrm{BR}}r_2e^{i\delta}
	\end{align*}
	
	These eight equations describe the two counter-propagating intracavity fields in each of subsection of the cavity model formed between the coupling element and a mirror. The notation $\mathrm{E}_{ijk}$ denotes the field in subsection $i\in(1,2)$ for mode $j\in(\mathrm{A},\mathrm{B})$ of propagating direction $k\in(\mathrm{L},\mathrm{R})$. In this work, mode A corresponds with $l=3$ and mode B corresponds with $l=0$. $\mathrm{E}_{0\mathrm{A}}$ is the input field. The parameter $\delta$ is a phase factors accrued by propagation: $\delta=\mathrm{kL}$ where L is the length of each cavity subsection (which we assume to be equal) and k is the wavenumber. If we vary $\delta$, we essentially varying the frequency of the imaginary laser probing this model cavity system. The transmitted fields are related to the intracavity fields by the transmission coefficient of the mirror through which one wishes to calculate transmission. The mirror through which we calculate the transmitted field is depicted in bold in Figure~\ref{fig:ImpMatch} with the corresponding intensity plotted below as a function of $\delta$. The calculated intensities are normalized to the input intensity.
	
	\begin{table}[ht]
		\renewcommand\thetable{S1}
		\centering
		\caption{Dual-wavelength cavity specifications \label{table:specs}}
		\begin{tabular}[t]{|p{0.35\linewidth}|p{0.275\linewidth}|p{0.275\linewidth}|}
			\hline
			& $780$~nm & $1529$~nm\\
			\hline
			Lower waist size & 19 $\mu$m & 27 $\mu$m\\ 
			Top 2x mirrors & 99.91$\%$ & 99.82$\%$ \\
			Bottom 2x mirrors & $>$99.9$\%$ (HR) & 99.94$\%$ \\
			Finesse $\mathcal{F}$ & 1900 & 1310 \\ 
			\hline
		\end{tabular}
	\end{table}
	
	For $r_1=r_2=\sqrt{0.9991}$ (the reflection coefficient of our top 2x cavity mirrors as listed in Table~\ref{table:specs}) and increasing $r$, the $l=0$ transmission increases from zero and saturates to $25\%$. For $r\sim0.001$, the calculated $l=0$ and $l=3$ transmission in Figure~\ref{fig:ImpMatch}b mimics the measured $l=0$ and $l=3$ transmission in Figure~\ref{fig:ImpMatch}a. In fact, the non-Lorentzian line shape of the $l=0$ mode is a result of the $l=0$ and $l=3$ coupling. Increasing $r$ couples the $l=0$ and $l=3$ more strongly, resulting in a vacuum Rabi-like splitting in both the $l=0$ and $l=3$ spectra. Thus, the non-Lorentzian line shape of $l=0$ is indicative of nonzero $l=0$ and $l=3$ coupling, but not enough to fully split the $l=0$ mode in the spectra.
	
	For $r_1=\sqrt{0.9991}, r_2=1$, and the same $r$ as in Figure~\ref{fig:ImpMatch}b, the $l=0$ mode is fully transmitted at $100\%$ as depicted in Figure~\ref{fig:ImpMatch}c. This change in $r_2$ is equivalent to making our double-ended cavity into single-ended cavity, where light can leak out of only one cavity mirror. Thus, applying the method presented in this paper to a low loss, single-ended cavity holds promise for achieving mode conversion at an external efficiency near $100\%$.
	
	\subsection{$\mathcal{E}_{3\rightarrow0}$ versus cavity-atom detuning}
	\label{SI:Det}
	
\begin{figure}
		\centering
		\includegraphics[width=1.0\linewidth]{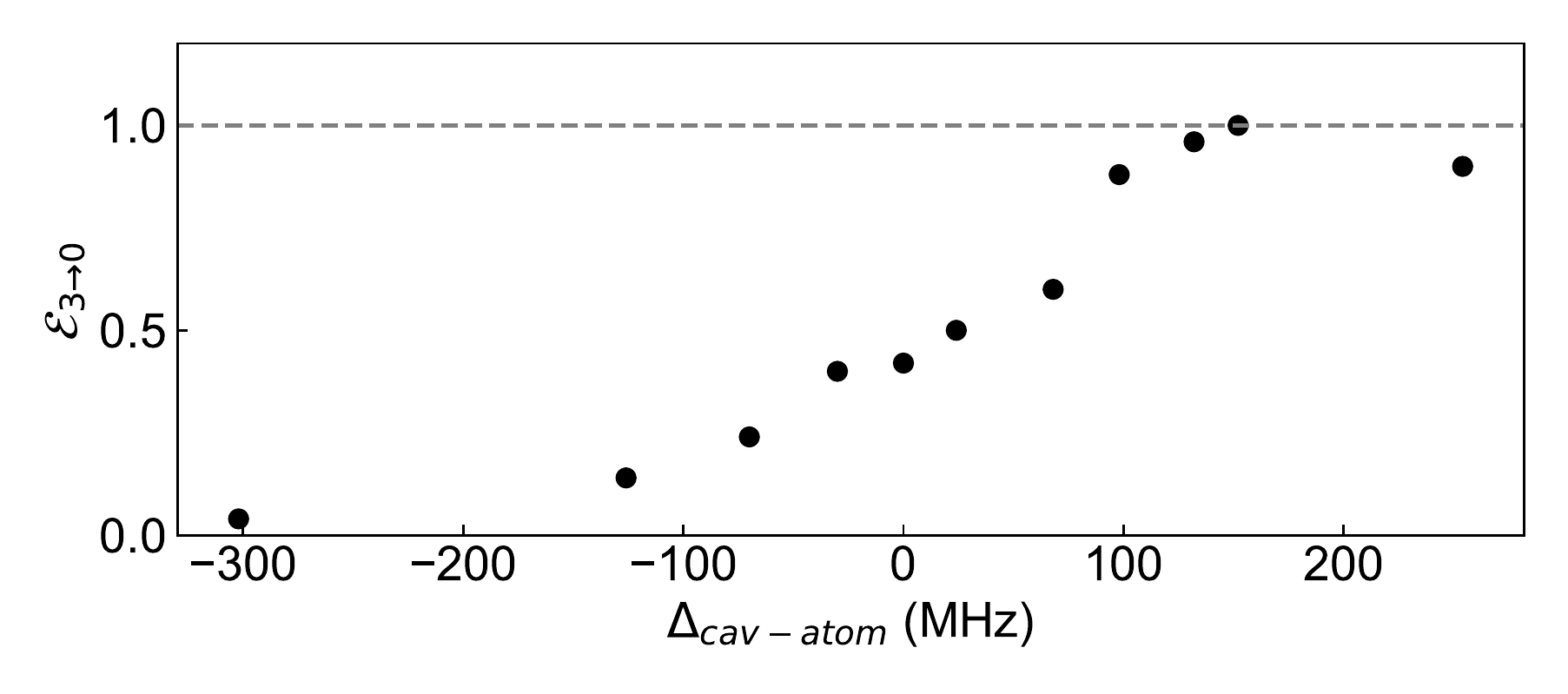}
		\caption{\textbf{Maximum conversion efficiency versus cavity-atom detuning.} In this work, the $l=3$ cavity mode was 130 MHz detuned from the $5S_{1/2}\rightarrow5P_{3/2}$ atomic transition, where we experimentally observed the highest $\mathcal{E}_{3\rightarrow0}$. All data was collected for the highest values of $\Omega$ and $N\eta$ used in this work.
			\label{fig:CavAtDet}}
	\end{figure}

	In this work, we operated in the dispersive regime, where the $l=3$ cavity mode was 130 MHz detuned from the $5\mathrm{P}_{3/2}$ state ($\Delta_{cav-atom}=130$ MHz). For the maximum values of $\Omega$ and $N\eta$ used in this work ($\Omega/(2\pi)=3.5$ GHz and $N\eta=560$), we experimentally observed high $\mathcal{E}_{3\rightarrow0}$ around this detuning as illustrated in Figure~\ref{fig:CavAtDet}. For smaller and opposite sign detunings, we observed a substantial decrease in $\mathcal{E}_{3\rightarrow0}$. While the source of this decrease has not yet been identified, we hypothesize it might arise from loss due to couplings to other twisted cavity modes or hyperfine levels of the $5\mathrm{P}_{3/2}$ state.
	
	\section{Theory}
	
	\subsection{Laguerre-Gaussian modes}
	
	The normalized electric field for a Laguerre-Gaussian mode $\mathrm{LG}_{lp}$ at the lower cavity waist $\omega$ is,
	
	\begin{align*}
		u_{lp}(r,\phi) =
		&\frac{C_{lp}}{w} \left( \frac{r\sqrt{2}}{w} \right)^{|l|}  L_{p}^{|l|}\left(\frac{2 r^{2}}{w^{2}}\right) \nonumber e^{-(r^{2}/w^{2})} e^{-i l \phi}
	\end{align*}
	
	\noindent
	where $r$ ($\phi$) is the radial (azimuthal) coordinate. The mode index $l$ describes the orbital angular momentum which manifests as a phase winding, whereas the mode index $p$ describes the number of radial, intensity `rings.' Both indices are integers with $p \geq 0$. $L_{p}^{l}\left(x\right)$ are the generalized Laguerre polynomials and the normalization constant $C_{lp} = \sqrt{\frac{2p!}{\pi\left( p+|l| \right)!}}$ to ensure $\bra{u_{lp}} u_{lp} \rangle = 1$.
	
	\subsection{Modeling conversion}
	\label{SI:Model}
	
	This section describes steps taken to model the conversion process in this work. We write down the full, time-dependent Hamiltonian then consider a simplified version of this Hamiltonian to computationally simplify spectra simulations. While the simulated spectra lack quantitative agreement with the experimental data, likely because the simplified Hamiltonian considers only a limited state space compared to the full Hamiltonian, they qualitatively capture main features of the data and are discussed here for the interested reader. The full, time-dependent Hamiltonian for the system described in this work written in the frame rotating with the $780$~nm probe laser of frequency $\omega_{\ell}$ is ($\hbar\equiv1$),
	
	\begin{align*}
		H(t) &= \sum_{n}^{N_{cav}} (\omega_n-\omega_{\ell}-i\frac{\kappa}{2})a_n^{\dagger}a_n \\
		& +\sum_{m}^{N_{at}} (\omega_p-\omega_{\ell}-i\frac{\Gamma_p}{2})p_m^{\dagger}p_m \\
		& +\sum_{m}^{N_{at}} (\omega_d-\omega_{\ell}-i\frac{\Gamma_d}{2})d_m^{\dagger}d_m \\
		& +\sum_{n}^{N_{cav}}\sum_{m}^{N_{at}} (g_{mn}p_m^{\dagger}a_n+g_{mn}^*p_m a_n^{\dagger}) \\
		& +\sum_{m}^{N_{at}} (\Omega_m(t)d_m^{\dagger}p_m+\Omega_m^*(t)d_m p_m^{\dagger}) \\
		& +\Omega_{\ell}(a_3^{\dagger}+a_3)
	\end{align*}
	
	\noindent
	where $\omega_n$ is the energy of the $n^{\text{th}}$ $780$~nm cavity mode, $\omega_p$ is the energy of the $5P_{3/2}$ state, $\omega_d$ is the energy of the $4D_{5/2}$ state, $\kappa$ is the cavity decay rate at $780$~nm, $\Gamma_p$ is the atomic decay rate of the $5P_{3/2}$ state, and $\Gamma_d$ is the atomic decay rate of the $4D_{5/2}$ state. The operators $a_n$, $p_m$, and $d_m$ annihilate a photon in the $n^{\text{th}}$ $780$~nm cavity mode, a $P$-state excitation for the $m^{\text{th}}$ atom, and a $D$-state excitation for the $m^{\text{th}}$ atom, respectively. The drive strength of the probe laser is represented by $\Omega_{\ell}$, which drives only the $l=3$ cavity mode. 
	
	The coupling strength $g_{mn}$, which couples the $P$-state of the $m^{\text{th}}$ atom and the $n^{\text{th}}$ $780$~nm cavity mode, can be expressed as
	
	\begin{align*}
		g_{mn} &= g_n u^{780}_{n_ln_p}(r_m,\phi_m) \\
	\end{align*}
	
	\noindent
	where $g_n$ is the single atom-photon coupling strength of the $n^{\text{th}}$ cavity mode and $u^{780}_{n_ln_p}(r_m,\phi_m)$ is the field of the $n^{\text{th}}$ $780$~nm cavity mode at the location of the $m^{\text{th}}$ atom. The $n^{\text{th}}$ $780$~nm cavity mode has $l$ index $n_l$ and $p$ index $n_p$. 
	
	\begin{figure*}
		\centering
		\includegraphics[width=1.0\linewidth]{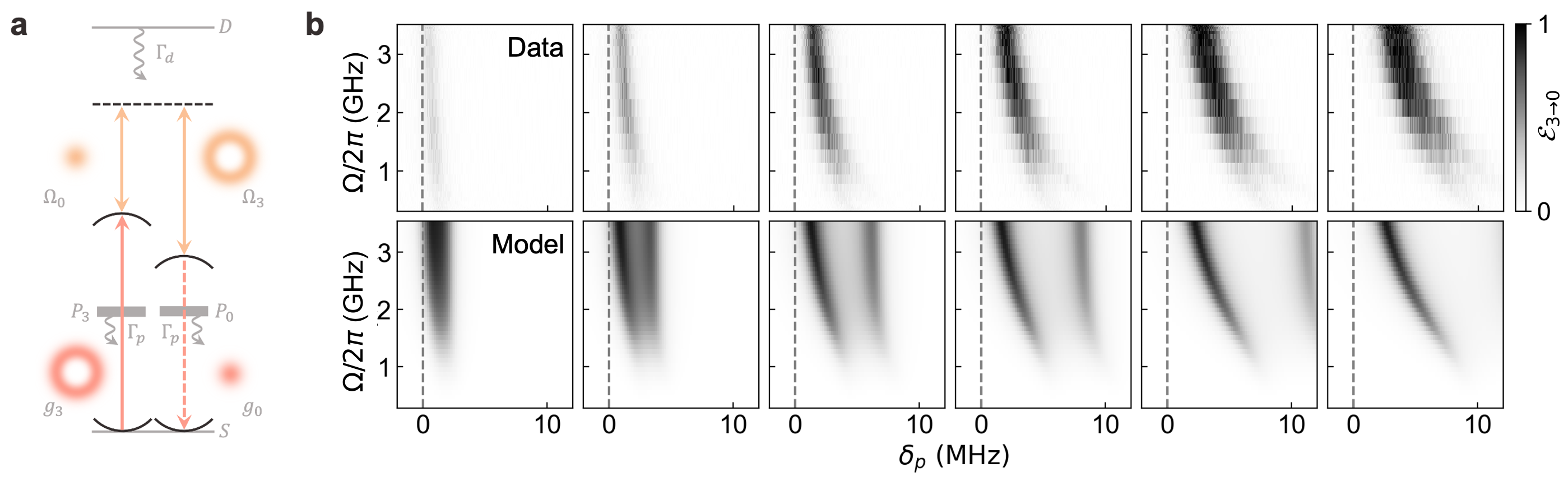}
		\caption{\textbf{Spectra predictions of a simple model.} A simplified collective state model of the conversion process described in the work is depicted in \textbf{a}. The $l=3$, $780$~nm mode couples to the collective $P_3$ state with effective coupling $g_3$, the $l=0$, $780$~nm mode couples to the collective $P_0$ state with effective coupling $g_0$, and both collective $P_3$ and $P_0$ states couple to a collective $D$ state with effective couplings $\Omega_3$ and $\Omega_0$, respectively. We solve for the expectation value of $a_0^{\dagger}a_0$, the transmission of the $l=0$, $780$~nm light, as a function of $\delta_p$, $\Omega$ (the $1529$~nm beam strength), and $N\eta$ using non-Hermitian perturbation theory where the drive term $\Omega_{\ell} (a_3 + a_3^\dagger)$ is the perturbation. The resulting, simulated spectra are plotted below the experimental spectra in \textbf{b}. All simulated parameters are identical to the experimental parameters, with the exception of $\Omega_3/\Omega_0=1.0$ instead of 1.7 as better agreement with the experimental data was observed. The simulated spectra has clear differences with the experimental spectra. Namely, the presence of an unobserved spectral feature and increased conversion at low $N\eta$. However, simulations at even lower $N\eta$ display an overall decrease in conversion akin to that in the data. Thus, we suspect this falloff in efficiency for lower $N\eta$ in simulations compared to experiment may be because this simple model excludes couplings to other collective states which act as loss channels. Despite the shortcomings of this model, it qualitatively predicts the saturation of $\mathcal{E}_{3\rightarrow0}$ to 1 for some minimum $\Omega$ and $N\eta$ and captures the shapes of the experimental spectra.
			\label{fig:ConvModel}}
	\end{figure*}
	
	The time-dependent coupling strength $\Omega_m(t)$, which couples the $D$-state of the $m^{\text{th}}$ atom and the $P$-state of the $m^{\text{th}}$ atom, can be expressed as
	
	\begin{align*}
		\Omega_m(t) &= \Omega_0 u^{1529}_{00}(r_m,\phi_m)\exp(i\omega_0^{1529}t) \\ 
		&+ \Omega_3 u^{1529}_{30}(r_m,\phi_m)\exp(i\omega_3^{1529}t)
	\end{align*}
	
	\noindent
	where $\Omega_0$ and $\Omega_3$ are coupling strengths dependent on the field strength of the $l=0$ component and $l=3$ component of the $1529$~nm beam, respectively. The frequencies of the $l=0$ component and $l=3$ component are $\omega_0^{1529}$ and $\omega_3^{1529}$, respectively. Ordinarily, the time dependence of the Hamiltonian due to $\Omega_m(t)$ can be eliminated by a transformation, but here the presence of dual frequencies $\omega_0^{1529}$ and $\omega_3^{1529}$ prevents this elimination. Instead, the time dependence must be handled with Floquet theory or by solving for the time dynamics of the system. Additionally, coupling terms are often simplified by assuming uniformity of electric fields across the atomic sample, but here this idea does not apply. For many atoms, simulating this system with time- and space-dependent terms can by quite slow. An alternative approach to simplifying the massive state space for many atoms is to work in the collective state picture after adiabatic elimination of the $4D_{5/2}$ state, though this process comes with its own challenges such as determining the couplings between collective states and identifying which collective states are the most meaningful.
	
	In light of these challenges, we considered a much simpler Hamiltonian to explore how well it could model the conversion spectra observed in this work. Fig.~\ref{fig:ConvModel}a depicts a modified level diagram described by the Hamiltonian,
	
	\begin{align*}
		H &= (\omega_3-\omega_{\ell}-\frac{i\kappa}{2})a_3^{\dagger}a_3 \\
		&+(\omega_0-\omega_{\ell}-\delta_{03}-\frac{i\kappa}{2})a_0^{\dagger}a_0 \\
		&+(\omega_p-\omega_{\ell}-\frac{i\Gamma_p}{2})P_3^{\dagger}P_3 \\
		&+(\omega_p-\omega_{\ell}-\delta_{03}-\frac{i\Gamma_p}{2})P_0^{\dagger}P_0 \\
		&+(\delta_{d3}-\omega_{\ell}-\frac{i\Gamma_d}{2})D^{\dagger}D \\
		&+g_0(a_0^{\dagger}P_0+a_0P_0^{\dagger})+g_3(a_3^{\dagger}P_3+a_3P_3^{\dagger}) \\
		&+\frac{\Omega_0}{2}(D^{\dagger}P_0+DP_0^{\dagger})+\frac{\Omega_3}{2}(D^{\dagger}P_3+DP_3^{\dagger}) \\
		&+\Omega_{\ell} (a_3 + a_3^\dagger) 
	\end{align*}
	
	\noindent
	which considers only $l=0$ and $l=3$ modes. Now, operators $P_0$, $P_3$, and $D$ annihilate collective excitations instead of excitations of a single atom, and $g_0$, $g_3$, $\Omega_0$, and $\Omega_3$ are effective couplings to collective states. The collective states corresponding with the $P_0$ and $P_3$ operators adopt the orthogonality properties of the LG modes and are each coupled through one of the $1529$~nm pathways to a collective $D$ state. Detunings $\delta_{03}$ and $\delta_{d3}$ are the frequency differences $\omega_0-\omega_3$ and $\omega_d-\omega^{1529}_{\ell}$, respectively, where $\omega^{1529}_{\ell}$ is the frequency of the $l=3$ component of the $1529$~nm beam.
	
	While this model falls short of quantitative agreement with the experimental data and predicts unobserved spectral features, it depicts the saturation of $\mathcal{E}_{3\rightarrow0}$ to 1 for some minimum threshold of $\Omega$ and $N\eta$ and captures the shapes of the experimental spectra (Fig.~\ref{SI:Model}b). Additional work is necessary to attain a better understanding of the minimum $\Omega$ and $N\eta$ needed to maximize the conversion efficiency and the conditions needed to suppress couplings to non-target LG modes, but the qualitative similarities between the modeled and experimental data provide some reassurance that the picture depicted in Fig.~\ref{fig:ConvModel}a is a step in the right direction.
	
\end{document}